# Optically Excited Two-Band Amplified Spontaneous Emission from a High-Current-Density Quantum-Dot LED

Namyoung Ahn[1,†], Young-Shin Park[1,2,†], Clément Livache[1], Jun Du[1,3], and Victor I. Klimov[1]*

[1]Nanotechnology and Advanced Spectroscopy Team, C-PCS, Chemistry Division, Los Alamos National Laboratory, Los Alamos, New Mexico 87545, USA

[2]Department of Chemical and Biomolecular Engineering, Korea Advanced Institute of Science and Technology, Daejeon 34141, Republic of Korea

[3]State Key Laboratory of Molecular Reaction Dynamics, Dalian Institute of Chemical Physics, Chinese Academy of Sciences, Dalian, Liaoning 116023, China

*klimov@lanl.gov

**Laser diodes based on solution-processable materials could benefit numerous technologies including integrated electronics and photonics, telecommunication, and medical diagnostics. An attractive system for implementing these devices is colloidal semiconductor quantum dots (QDs). The primary challenge that hampered progress towards a QD laser diode (QLD) has been fast nonradiative Auger decay of optical-gain-active multicarrier states. Recently, this problem has been resolved by employing continuously graded QDs (cg-QDs) wherein Auger recombination is strongly suppressed. The use of these structures allowed for demonstrations of optical gain with electrical pumping and optically-excited lasing in multilayered LED-like devices. Here we report on achieving the next critical milestone towards a QLD, which is the demonstration of optically excited amplified spontaneous emission from a fully functional high-current density electroluminescent device. This advance has become possible due to excellent optical gain properties of novel 'compact' cg-QDs and a new LED architecture, which allows for concerted optimization of its optical and electrical properties. The results**

**of this work strongly suggest the feasibility of the final step towards a functional QLD, which is the demonstration of lasing with electrical pumping.**

Semiconductor laser diodes based on solution-processable materials have been pursued across multiple fields including organic semiconductors,[2] perovskites,[5] and colloidal quantum dots (QDs).[1,3] Their development has been motivated by prospective applications in lighting,[6] communications,[7] sensing,[8] and data storage.[9] Colloidal QDs are promising materials for realizing these devices. In addition to being compatible with standard solution-based fabrication and processing techniques, they offer facile spectral tunability, achievable by size and/or composition control, and high (near-unity) emission quantum yields.[10] Further, a wide separation between their quantized states prevents thermal depopulation of the band-edge emitting levels and thereby reduces lasing thresholds and improves thermal stability of lasing characteristics.[11]

The primary difficulty for realizing lasing with colloidal QDs is fast, confinement-enhanced nonradiative Auger recombination wherein the recombination energy of an electron-hole (e-h) pair (exciton) is not emitted as a photon but dissipates *via* Coulomb energy transfer to the third carrier.[12,13] This complication arises from a multifold degeneracy of the QD band-edge levels, because of which the development of optical gain requires that QDs are populated with more than one exciton per dot on average. As a result, optical gain relaxation is controlled by multiexciton Auger recombination which leads to very short (typically sub-100 ps) gain lifetimes ($\tau_\text{g}$).[14] This represents an especially serious problem in the case of continuous-wave optical or direct-current electrical pumping.[1,3] In particular, it leads to extremely high lasing thresholds that are often well above damage thresholds of colloidal QDs and/or other materials (*e.g.*, organic molecules) used in solution-processed devices.[15-17]

Recently, strong suppression of Auger decay has been achieved by employing continuously graded QDs (cg-QDs) wherein a CdSe core is enclosed into a $Cd_xZn_{1-x}Se$ shell with $x$ varied from 1 to 0 along the radial direction.[4] For improved stability, the $CdSe/Cd_xZn_{1-x}Se$ cg-QDs are usually overcoated with a thin protective layer of $ZnSe_yS_{1-y}$.[18] The invention of cg-QDs allowed for several advances of direct relevance to the development QD laser diodes (QLDs). These include the demonstration of band-edge (1S) optical gain with d.c. electrical pumping[4] and, more recently, the realization of pulsed QD devices capable of stable operation at current densities ($j$) of ~1,000 A cm$^{-2}$, which was sufficient to achieve optical gain for both the 1S and the higher-energy 1P transition.[3]

Another important milestone towards a functional QLD has been the realization of optically excited lasing in LED-like devices with an integrated distributed feedback (DFB) resonator.[19] The reported structures featured a multilayered stack of a standard 'inverted' QD LED which lacked only a top hole-injecting electrode. Despite a fairly small thickness of an active cg-QD layer (~50 nm) and the presence of 'lossy' charge-transport layers, these devices showed good lasing performance under pulsed optical excitation. They also exhibited bright electroluminescence (EL) upon addition of a final $MoO_x$/Al contact. However, this last step resulted in suppression of lasing, indicating that the overall optical losses in the complete EL device were greater than modal gain generated by the QDs.

Here we resolve the 'quenching problem' by implementing an integrated approach in which optical gain/loss engineering is combined with engineering of charge injection so as the resulting devices exhibit strong gain, suppressed optical losses, excellent waveguiding properties, and high current densities. Using this approach, we are able to achieve maximal (saturated) values of 1S optical gain with electrical pumping and also realize optically-excited two-band (1S and 1P) amplified spontaneous emission (ASE) in the same electroluminescence (EL) active device. This indicates that both the 1S and 1P modal gain is sufficiently high to overcome optical losses and allow for efficient amplification of light guided along the QD active layer.

**Compact continuously graded QDs as optical gain materials**

The modal gain realized in an LED device stack ($G_{mod}$) can be presented as the product of a mode confinement factor ($\Gamma_{QD}$) of an active QD layer and the QD 'material gain' defined as the optical gain of an infinitely thick QD solid ($G_{mat}$): $G_{mod} = \Gamma_{QD} G_{mat}$. Here, we aim to enhance both $G_{mat}$ and $\Gamma_{QD}$ via, respectively, enhancing gain performance of the QD medium and optimizing the overall structure of the device stack.

First, we discuss our approach to enhancing $G_{mat}$. This quantity can be presented as $G_{mat} = \sigma_{gain} n_{QD}$, where $\sigma_{gain}$ is the QD gain cross-section and $n_{QD}$ is the QD density in the film. For our type-I cg-QDs, the oscillator strength of the band-edge transition, which defines $\sigma_{gain}$, is expected to be virtually independent on QD dimensions.[20,21] Therefore, we seek to boost $G_{mat}$, by increasing QD packing density ($n_{QD}$). In the case of close-packed solids, $n_{QD}$ scales inversely with QD volume ($V_{QD}$) and, hence, it can be increased by decreasing the overall QD size. To test the validity of this approach, we synthesize a series of CdSe/Cd$_x$Zn$_{1-x}$Se/ZnSe$_{0.5}$S$_{0.5}$/ZnS QDs (see Methods and Supplementary Fig. 1 for details of the synthesis) with a similar core radius (*ca*. 2.6 nm) and a varied thickness of the graded layer (Fig. 1a). Then, we process them into a 300-nm thick close-packed film on top of a glass slide and characterize their optical gain using a variable stripe length (VSL) technique (Methods).

The VSL measurements reveal the anticipated trend, which is the increase in $G_{mat}$ with decreasing QD dimensions (Fig. 1b). The measured $G_{mat}$ scales approximately as $1/V_{QD}$ (Fig. 1c), as expected for direct proportionality between $G_{mat}$ and the QD packing density. In particular, for the original cg-QDs with $R_{QD}$ = 9.4 nm (ref[22]), the band-edge (1S) gain coefficient is ~200 cm$^{-1}$. It is boosted more than three-fold (to 650 – 800 cm$^{-1}$) for smaller-size, ~6-nm cg-QDs, which is consistent with the decrease of $V_Q$ by a factor of 3.8. Importantly, despite their reduced dimensions, the 'compact' cg-QDs (hereafter referred to as ccg-QDs) exhibit strong suppression of Auger recombination. Specifically, their biexciton lifetimes ($\tau_{XX}$ of 0.9 – 1.3 ns; Fig. 1b and Supplementary Fig. 2) are similar to those of the original cg-QDs with a thick graded layer.[22]

Because of their strong optical-gain performance, we use the ccg-QDs in the present work. Specifically, for our device-related studies, we prepare samples whose core radius is 2.6 nm and the overall radius is 6.25 nm (Fig. 1a). In Fig. 1d (left), we present their photoluminescence (PL; red) and linear absorption ($\alpha_0$; black) spectra along with its second derivative ($\alpha_0''$; blue). As the original cg-QDs,[22] the ccg-QDs show an enhanced splitting (56 meV) between light and heavy hole states (abbreviated as 'lh' and 'hh'; Fig. 1d, right), ascribed previously to asymmetric compression of the CdSe core.[22] The lh-hh splitting manifests as a double-peak band-edge (1S) feature observed in both $\alpha_0$ and $\alpha_0''$ spectra. These spectra also exhibit the third, higher-energy feature due to the transition involving the 1P electron and hole states (Fig. 1d, right).

**Optimization of optical gain/loss characteristics *via* device stack engineering**

In the next step, we aim to reduce optical losses and simultaneously increase $\Gamma_{QD}$ *via* the optimization of individual elements of the LED device stack and its overall structure. A common feature of a traditional 'inverted' QD LED[23] is a bottom electron-injecting electrode (cathode) made of indium tin oxide (ITO) used for its optical transparency and high electrical conductivity. However, standard ITO may inhibit light amplification because of strong free-carrier absorption in the range of visible wavelengths. Further, due to its high refractive index ($n_{ITO}$ = 1.89 at $\lambda$ = 600 nm, ref[24]), which is comparable to that of the QD solid ($n_{QD}$ = 1.92, ref[19]), the ITO layer tends to 'pull' the optical field from the active QD layer, which reduces $\Gamma_{QD}$ and thereby diminishes modal gain. These problems were identified in ref[19] and were successfully tackled using so-called low-index ITO or L-ITO (1:2 mixture of ITO and SiO$_2$). This material exhibits a considerably lower absorption coefficient compared to standard ITO (160 cm$^{-1}$ versus 483 cm$^{-1}$ at $\lambda$ = 600 nm; Fig. 2a) and features a reduced refractive index (1.6 versus 1.89). Based on these favorable characteristics, we use L-ITO in our devices as a cathode.

Another common element of standard inverted QD LEDs is an electron transport layer (ETL) made of n-type colloidal or sol-gel ZnO.[19,25] The Fermi level of this material matches well the conduction band of CdSe QDs, which facilitates electron injection.[26] However, the ZnO refractive index ($n_{ZnO}$ = 1.95, ref[27]) is higher than that of the QD solid which reduces $\Gamma_{QD}$. In addition, the ZnO layer is optically lossy[28] and its low thermal conductivity can hamper heat exchange at high current densities leading to device breakdown (Supplementary Fig.3). To avoid these detriments, we exclude the ZnO ETL from our device stack and attempt to inject electrons into the QDs directly from the L-ITO contact. As discussed later in this work, this approach indeed improves waveguiding properties of our devices and, importantly, still allows us to reach very high current densities sufficient for generating strong optical gain.

Additionally, we also modify the hole injection/transport part of the LED. Commonly, hole injection is accomplished using a combination of an organic hole-transport layer (HTL) made of, for example, tris(4-carbazoyl-9-ylphenyl)amine) (TCTA, 50 nm), followed by a thin MoO$_x$ (10 nm) hole injection layer (HIL). (Fig. 2b, top) [4,18] From the QLD standpoint, the detriment of this design is extremely high optical losses in MoO$_x$ ($\alpha_{MoOx}$ = 3153 cm$^{-1}$ at 600 nm, ref[29]), which results in undesired quenching of waveguided modes.[19] Strikingly, when MoO$_x$ is combined with TCTA, the absorbance of the resulting TCTA/MoO$_x$ bi-layer becomes even greater than the sum of $\alpha_{MoOx}$ and $\alpha_\text{TCTA}$ (Fig. 2b, top), leading to additional optical losses. This likely occurs due to the increased free-carrier density in the organic HTL caused by the inflow of holes from MoO$_\text{x}$.[30]

To mitigate these problems, we exclude MoO$_\text{x}$ from our devices and instead implement hole injection using an all-organic bi-layer made of 2,2',7,7'-tetrakis[N-naphthalenyl(phenyl)-amino]-9,9-spirobifluorene (Spiro-2NPB) and dipyrazino[2,3-f:2',3'-h]quinoxaline-2,3,6,7,10,11-hexacarbonitrile (HAT-CN).[31] The absorbance of individual layers in this combination is much lower than that of MoO$_\text{x}$ and, importantly, it is not increased due to interfacial effects when they are joined together (Fig. 2b, bottom).

To compare optical properties of our new device stack with those of the standard inverted LED, we conduct their modeling using a finite element method implemented with a standard COMSOL software (Methods). The structures of the simulated devices, completed with a top contact (anode) made of silver, are ITO/ZnO/QDs/TCTA/MoO$_\text{x}$/Ag (Fig. 2c) and L-ITO/QDs/Spiro-2NPB/HAT-CN/Ag (Fig. 2d). Importantly, the presence of the metal anode, leads to strong quenching of transverse magnetic (TM) modes (Supplementary Fig. 4). Therefore, in our modeling, we focus on the fundamental transverse electric (TE$_0$) mode.

The results of simulations are displayed in Fig. 2c,d. They include profiles of transverse electric field (black line) and optical power losses (red line), as well as the spatial distribution of the electric field along the waveguide formed by the QD layer (red/blue shading) at $\lambda$ = 600 nm. In the structure with a traditional inverted LED design (Fig. 2c), the presence of the ITO layer leads to considerable optical losses characterized by the effective loss coefficient ($\alpha_\text{ITO}$) of 220 cm$^{-1}$. Other device elements yield a small additional contribution of 42.5 cm$^{-1}$. As a result, the total loss coefficient ($\alpha_\text{total}$) is 262.5 cm$^{-1}$ (Fig. 2e, blue circles).

In addition to large losses, a further problem of the standard LED architecture is a nonoptimal optical field profile which peaks not in the QD part of the device, but at the interface of the ZnO and ITO layers (Fig. 2c, black line), as a consequence of their high refractive indices. This reduces the confinement factor for the QD layer (Fig. 2f) and leads to high threshold values of the QD material gain ($G_{mat,th}$) required for lasing. In particular, for the structure in Fig. 2c, $\Gamma_{QD} = 0.21$, which yields $G_{mat,th} = \alpha_{total}/\Gamma_{QD} = 1250$ cm$^{-1}$. Since the maximal 1S gain achievable with ccg-QDs is of ~800 cm$^{-1}$ (Fig. 1b,c), the traditional LED design is not suitable for realizing a QLD.

The newly proposed LED design (Fig. 2d) is much more favorable for light amplification. In this case, the QD film is sandwiched between the L-ITO cathode and the organic HTL whose refractive indices are lower than that of the QDs ($n_{L-ITO} = 1.6$, $n_{HTL} = 1.8$). As a result, the peak of the optical field distribution shifts into the QD layer (Fig. 2d, black line), which leads to about two-fold increase of the mode confinement factor ($\Gamma_{QD} = 0.4$; Fig. 2e). Further, thanks to the reduced absorbance of L-ITO, the overall optical loss drops to ~143 cm$^{-1}$ (Fig. 2d, red line and Fig. 2e, green diamonds). As a result of these improvements, $G_{mat,th}$ is reduced to 358 cm$^{-1}$, a value easily accessible with ccg-QDs (Fig. 1b,c).

**EL properties of LEDs with optimized optical-gain characteristics**

To practically implement the developed design, we fabricate LEDs whose structure and schematic band diagram are displayed in Fig. 3a (left and right panels, respectively); see Methods for fabrication details. In particular, we spin-coat 6-monolayers of ccg-QDs on top of an L-ITO substrate and then use thermal evaporation to deposit the spiro-2NPB HTL, the insulating LiF layer with a 50-μm-wide gap, the HAT-CN HIL, and the Ag anode prepared as a narrow (300-μm wide) strip orthogonal to the gap in the LiF membrane (see top-view and cross-section scanning electron microscopy (SEM) images in Fig. 3b; left top and bottom, respectively). A combination of the specially-shaped Ag contact and the aperture in the insulting LiF layer leads to two-dimensional confinement of the injection area (Fig. 3b, right), commonly described as 'current focusing', which helps improve heat exchange of the emitting volume with the environment and thereby boost the maximal current density ($j$) achievable before device breakdown.[3]

Direct contact of the ccg-QDs with the conductive L-ITO cathode could, in principle, lead to emission quenching due to nonradiative pathways associated with charge and/or energy transfer. However, time-resolved PL measurements indicate that PL dynamics of the ccg-QDs prepared on top of the L-ITO substrate is virtually identical to that of dots on the glass substrate, indicating the absence of additional nonradiative decay channels (Supplementary Fig. 5). This is likely due to the shielding effect of the graded $Cd_xZn_{1-x}Se$ layer and the $ZnSe_{0.5}S_{0.5}$/ZnS shell that isolate the emitting CdSe core from the electrode.[18]

Next, we examine electro-optical characteristics of the fabricated LEDs under both room temperature ($T$ = 300 K) and cryogenic cooling with liquid nitrogen (LN, $T$ = 80 K). To reduce overheating of the devices at high $j$, we drive them using short-pulse excitation (0.5 μs pulse duration, 100 Hz repetition rate).[3] In Fig. 3c, we present the current-density-voltage ($j$-$V$) characteristic (black line) and the $V$-dependent EL intensity (blue dashed line) of the LN-cooled LED. As $V$ is increased, the $j$-$V$ curve features a transition from ohmic conductance ($j \propto V^1$) to charge transport in the 'trap-filled limit' ($j \propto V^4$).[32] During the second phase, the current density rapidly increases reaching $j$ of 361 A cm$^{-2}$, which is well above the optical-gain threshold.[1,3] The realization of such high values of $j$ is facilitated by the implemented 'current-focusing' LED design and the use of pulsed bias, the measures applied previously to reduce the amount of generated heat and to improve heat outflow from the active device volume.[3,33,34] Due to high $j$, the device produces intense EL at $j$ = 169 A cm$^{-2}$ (Fig. 3c, blue dashed line). At higher $j$, $L$ saturates and then exhibits a slight drop. This behavior, however, is reversible, indicating that it is not due to device degradation but rather due to temperature-induced EL quenching.[35]

The developed devices also exhibit strong performance at room temperature. In particular, they reach current densities up to 557 A cm$^{-2}$ (Supplementary Fig. 6a), which exceeds maximal $j$ obtained at LN temperature. Due to higher $j$, we are able to achieve the regime of trap saturation marked by the transition to the $j \propto V^2$ dependence (observed at $V$ > 20 V), which is typical of 'trap-free space-charge-limited' conductance.[32]

Due to high current densities accessible with our devices, they allow us to realize the unusual regime of multiband EL. In Fig. 3d, we present spectrally resolved EL recorded at LN temperature for increasing current density. At $j$ = 15 A cm$^{-2}$, we observed a standard single-band EL due to the band-edge transition, which couples the 1S$_e$ electron and 1S$_{hh}$ heavy hole states (Fig. 3d, left). The spectrum recorded at $j$ = 68.1 A cm$^{-2}$ exhibits a pronounced broadening on its higher-energy side, which is due to the emergence of emission involving the 1S$_{lh}$ state (Fig. 3d, middle). The corresponding band obtained via a two-band Lorentzian fit (shown by orange shading) is separated from the band-edge feature (pink shading) by 57 meV. This is consistent with the light-heavy hole splitting observed in the absorption spectrum (Fig. 1d).[4] The two-band EL structure also manifests in the 2$^{nd}$ derivative of the EL signal (blue line in Fig. 3d, middle), which exhibits two peaks separated by 57 meV.

At $j$ = 68.1 Acm$^{-2}$, the EL spectrum exhibits further broadening to higher energies indicating the emergence of the third band (Fig. 3d, right). It is well pronounced in the 2$^{nd}$ derivative of the EL spectrum, based on which we infer that it is separated from the band-edge feature by 125 meV. This is consistent with the spacing between the 1S$_e$ –1S$_{hh}$ and the 1P$_e$ –1P$_{hh}$ transitions observed in the linear absorption spectrum (Fig. 1d), suggesting that the third EL band is due carriers occupying the 1P electron and hole states.

These observations can be explained by the progressive filling of higher energy states with increasing QD occupancy. If the number of the e-h pairs per dot ($N_{eh}$) is 2 or less, all injected carriers reside in the 2-fold-degenerate $1S_e$ and $1S_{hh}$ states, which leads to single-band EL (inset of Fig. 3d, left). At higher occupancies, the carriers are forced into higher energy $1P_e$ and $1S_{lh}$ states, which leads to the emergence of the $1S_e$ –$1S_{hh}$ feature (inset of Fig. 3d, middle). When $N_{eh}$ exceeds 4, the holes start to fill the $1P_{hh}$ level, which opens an additional radiative channel due to the $1P_e$ –$1P_{hh}$ transition (inset of Fig. 3d, right). This analysis suggests that the current densities realized in our devices are sufficient to completely fill the $1S_e$ and $1S_{hh}$ states, which requires $N_{eh}$ = 2. Further, since at the highest $j$, the $1S_e$ –$1S_{lh}$ feature also reaches saturation, this implies that the number of e-h pairs per dot exceeds 4. These results indicate that the developed LEDs allow us to achieve saturation of the band-edge gain (occurs when $N_{eh} \geq 2$, ref[1,3]) and, as a result, realize gain values comparable to those obtained with short-pulse optical pumping (Fig. 1b,c).

We are also able to achieve high per-dot occupation factors at room temperature (Supplementary Fig. 6b,c). In particular, at the highest current densities, we observe the intense $1P_e$ –$1P_{hh}$ feature indicating that $N_{eh}$ exceeds 4. In fact, in this case, the realized values of $N_{eh}$ are higher than those at LN temperature as indicated by the greater ratio of the amplitudes of the 1P and 1S features (0.6 versus 0.3; Supplementary Fig. 7).

**Optically excited ASE from fully stacked EL-active devices**

Next, we demonstrate that in addition to delivering strong optical gain, the developed high-$j$ LEDs are capable to produce optically excited ASE in a complete, EL-active device. In these experiments, the EL-active part of the device is excited through the transparent ITO electrode using a femtosecond pulsed laser (<190-fs pulse duration, 343-nm excitation wavelength) whose beam is tightly focused so as to illuminate QDs exclusively within the injection area defined by the 'current-focusing' elements of the device (Fig. 3b, top and Fig. 4a). This type of excitation allows us to re-create the situation of electrical pumping and thereby accurately evaluate the effect of losses arising from all elements of a fully stacked LED.

In Fig. 4b, we display pump-intensity dependent PL spectra collected from the cleaved edge of the device cooled to LN temperature. As per-pulse pump fluence ($w_p$) is increased, we observe a clear transition from spontaneous emission to ASE, first at the band-edge 1S transition and then at the higher-energy 1P transition. The development of ASE is evident in both a sharp growth of the PL peak amplitude ($I_{PL}$, Fig. 4c, open symbols) and pronounced line narrowing (Fig. 4c, solid symbols). In particular, the emergence of ASE is accompanied by the change in the log-log slope of the $I_{PL}$-vs.-$w_p$ dependence from 0.86 to 1.57 for the 1S feature (Fig. 4c, left), and from 1.54 to 5.51 for the 1P band (Fig. 4c, right). The ASE polarization corresponds to the TE waveguided modes (Supplementary Fig. 8). This is in agreement with our modeling which predicts strong quenching of the TM modes (Supplementary Fig. 4).

Based on the conducted measurements, the 1S and 1P ASE thresholds are $w_{ASE,1S}$ = 56.4 and $w_{ASE,1P}$ = 88.2 µJ cm$^{-2}$. If defined in terms of the average QD occupancy, $\langle N_{eh} \rangle$, they translate into $\langle N_{eh} \rangle \approx 4.5$ and 7, respectively. These two values are higher than the 'ideal' optical gain thresholds (1 and 6 e-h pairs per dot; Supplementary Note 1), which is a result of optical losses arising from various elements of the LED. Importantly, however, in the developed devices, the modal gain is sufficiently high to overcome optical losses, which leads to the desired effect of light amplification. This is a direct result of the conducted optimization of the optical field profile across the device stack, improved optical-gain properties of the ccg-QDs, and elimination (or replacement) of lossy device components based on ZnO, $MoO_x$, and standard ITO.

To analyze the interplay between optical gain and optical losses, we investigate ASE characteristics of partial (incomplete) device stacks that comprise selected sub-units of a full LED (Figs. 4d-f and Supplementary Fig. 9; $T$ = 80 K). In the electrode-free structure, which contains ccg-QDs sandwiched between the glass substrate and the organic HTL/HIL (Fig. 4d), ASE develops at low pump fluences of ~6 µJ $cm^{-2}$ (1S) and ~28 µJ $cm^{-2}$ (1P), due to the lack of appreciable optical losses in either the underlying substrate or the top organic layers (Fig. 2b).

In the device with the L-ITO layer inserted between the QDs and the glass substrate (Fig. 4e), the ASE thresholds remain virtually unchanged (8.8 and 28.2 µJ $cm^{-2}$ for the 1S and 1P features, respectively). The situation is different in the device wherein instead of the bottom L-ITO layer, we deposit a top Ag electrode (Fig. 4f). In this case, the ASE thresholds exhibit a considerable (about two-fold) increase (to 21.2 and 56.4 µJ $cm^{-2}$, for the 1S and 1P bands, respectively), observed simultaneously with the decrease of the log-log slope characterizing the ASE growth. These observations indicate that the metal contact is a dominant source of optical losses in our 'waveguiding' LEDs, while the specially engineered L-ITO electrode is responsible only for a small fraction of the overall device loss.

**Quantitative optical gain/loss analysis**

Here we quantify modal gain and overall optical losses in our LEDs using a combination of VSL measurements and simulations (Supplementary Fig. 10). First, we apply a VSL method to a ccg-QD layer within a fully assembled LED. In this case, the active area was extended to 1.5 × 1.5 $mm^2$ to allow for a wider range of stripe lengths. These measurements yield the net gain of the device ($G_{net} = G_{mod} - \alpha_{loss}$). Figure 5a presents a scheme of the measurements (inset) as well as a progression of the PL spectra for the increasing stripe length. Based on these measurements, at the highest pump fluence ($w_p$ = 248 µJ cm$^{-2}$), the net gain reaches ~5 cm$^{-1}$ (1S) and 220 cm$^{-1}$ (1P). (Fig. 5b,c). The net gain coefficients for other pump levels are displayed in Supplementary Fig. 11a,b and shown by green squares in Fig. 5d and 5e for the 1S and 1P bands, respectively.

Next, we infer the material gain by conducting pump-fluence-dependent VSL measurements on a thick ccg-QD film assembled on top of a low-loss glass substrate (Supplementary Fig. 11c,d). Based on the measurements and the computed mode confinement factor ($\Gamma_{QD}$ = 0.92), we obtain the material gain ($G_{mat}$) of our ccg-QDs plotted in Fig. 5d (1S) and 5e (1P). These data indicate a quick growth of $G_{mat}$ with increasing $w_p$, followed by saturation. The saturated material gain coefficients are 543 cm$^{-1}$ (1S) and 1,071 cm$^{-1}$ (1P). Using these results and $\Gamma_{QD}$ computed for our devices (Fig. 2f), we obtain the $w_p$-dependent 1S and 1P modal gain coefficients $G_{mod}$ (Fig. 5d,e; blue diamonds and blue dashed lines). Based on these data, the maximal modal gain realized in our devices is 190 cm$^{-1}$ for the 1S transition and 407 cm$^{-1}$ for the 1P transition. Using these values along with previously derived net gain coefficients in the device, we obtain that the loss coefficients ($\alpha_{loss} = G_{mod} - G_{net}$) are 185 cm$^{-1}$ (1S) and 187 cm$^{-1}$ (1P) bands. (Fig. 5d,e and Table 1) These values are in close agreement with our simulations (Fig. 2e), according to which $\alpha_{loss}$ is around 150 cm$^{-1}$ for both the 1S and 1P transitions. The higher value of $\alpha_{loss}$ obtained from the measurements (~40 cm$^{-1}$ difference) is likely due to additional losses arising from light scattering, not accounted for in the modeling.

While our devices exhibit ASE at $T$ = 80 K, the ASE is quenched at room temperature (Supplementary Fig. 12). Since ccg-QD gain is virtually the same at $T$ = 80 and 300 K (Supplementary Fig. 13), the observed quenching likely occurs due to temperature-induced increase in optical losses. In particular, the absorbance of a Ag layer is known to increase with temperature as a result of effects of electron-phonon scattering.[36] The increase in temperature can also raise effective doping levels of the L-ITO electrode and the organic layers and thereby boost free-carrier absorption.[37]

Although our devices are capable to generate large modal gain with electrical pumping, they do not exhibit electrically excited ASE, which is likely due to considerable overheating at high current densities. Indeed, although in the case of cryogenic colling, the nominal temperature of our LEDs is 80 K, the actual temperature can be much higher due to heat accumulation at high $j$.[3] This effect is evident in the progressive red shift of the EL spectrum with increasing $j$ (Supplementary Fig. 6b). Based on the magnitude of the observed shift (Supplementary Fig. 14), the device temperature near breakdown reaches 180 K, which is 100 K higher than the nominal temperature. As discussed earlier, the increase in the temperature leads to increased optical losses which apparently become so high that they overwhelm optical gain.

**Conclusions**

In conclusion, we have developed a new class of QD LEDs ('waveguiding LEDs') that combine two important functionalities of direct relevance to ongoing efforts of the development of QLDs. In particular, these devices are capable of generating extremely high current densities of up to ~ 560 A cm$^{-2}$, which are sufficient for realizing saturation of the 1S optical gain and even achieving partial population of the higher-energy 1P state. The same devices also exhibit both 1S and 1P ASE under optical pumping indicating that their optical gain overwhelms optical losses despite the presence of conducive layers required for efficient charge injection. This advance is a result of the integrated approach that combines optical gain/loss engineering with optimization of charge injection. In particular, the use of novel ccg-QDs allows us to boost material gain and simultaneously achieve strong suppression of Auger decay. Further, we engineer a cross-section profile of the refractive index so as to boost the mode confinement factor for the active QD layer. In addition, we reduce optical losses by removing (or modifying) strongly absorbing charge-transport/injection layers such as the ITO cathode (replaced with the L-ITO layer), a ZnO ETL (eliminated completely) and the TCTA/MoOx HTL/HIL (replaced with the all-organic HAT-CN/Spiro-2NPB HTL/HIL). As a result of these concerted efforts, we realize LEDs wherein the QD layer acts as an efficient waveguide amplifier with the large net optical gain of >200 cm$^{-1}$. Importantly, the conducted modifications preserve good electrical characteristics which allow for obtaining ultra-high current densities sufficient for realizing strong (saturated) optical gain. The remaining challenge is the realization of ASE/lasing with electrical pumping. Based on the conducted analysis, this would require further optimization of the LED structure for suppressing device overheating which presently leads to strong increase in optical losses under high current densities. This could be accomplished through, for example, the reduction of device serial resistance (*e.g.*, *via* introduction of conductive inter-dot 'linkers') and/or improved heat management.

# Methods

***Materials*** Cadmium acetate dihydrate ($Cd(OAc)_2 \cdot 2H_2O$, 98%, Aldrich), zinc acetate ($Zn(OAc)_2$, 99.99%), oleic acid (OA, 90%, Alfa Aesar), 1-octadecene (ODE, 90%, Aldrich), trioctylphosphine (TOP, 97%, Strem), sulfur (99.999%, Alfa Aesar), selenium (shot, 2-6 mm, 99.998%, Alfa Aesar), chloroform (99.9%, Fisher), ethanol (EtOH, Alfa Aesar), toluene (anhydrous, 99.8%, Aldrich), hexane (anhydrous, 95%, Aldrich), 1-propanol (anhydrous, 99.7%, Aldrich), and octane (anhydrous, 99%, Aldrich) were used as received. 2,2',7,7'-Tetrakis[N-naphthalenyl(phenyl)-amino]-9,9-spirobifluorene (Spiro-2NPB) and Dipyrazino[2,3-f :2',3'-h]quinoxaline-2,3,6,7,10,11-hexacarbonitrile (HAT-CN) were purchased from Lumtec. Silver pellets (Ag, 99.99%) were purchased from Kurt J. Lesker.

## Synthesis of ccg-QDs

***Precursors.*** 0.5 M cadmium oleate solution was prepared by dissolving 10 mmol $Cd(OAc)_2 \cdot 2H_2O$ in 10 mL OA and 10 mL ODE. The mixture was loaded into a three-neck flask and heated at 120 °C under vacuum for 1 hour. When the solution turned clear, the atmosphere was switched to nitrogen and the solution was kept at 100 °C for further use. 0.5 M $Zn(OA)_2$ solution was prepared by mixing 20 mmol $Zn(OAc)_2 \cdot 2H_2O$, 20 mL OA, and 20 mL ODE in a three-neck flask and degassing the mixture at room temperature for 5 min. Then, the solution was placed under nitrogen atmosphere and heated to 130 °C for 1 hour. After the solution turned clear, it was again carefully degassed under vacuum at 130 °C for 1 hour. The solution was then kept under nitrogen at 120 °C until further use. 2 M TOP-Se solution was prepared by stirring 40 mmol selenium in 20 mL TOP in a glove box until the complete dissolution of Se (typically, overnight). 2 M TOP-S was obtained by stirring 40 mmol sulfur in 20 mL TOP in a glove box until complete dissolution.

***Preparation of solutions for continuous injection.*** The ccg-QD synthesis was conducted *via* continuous injection of four precursor solutions. Solution (A) was obtained by mixing 0.5 mL Cd-oleate (0.5 M), 0.125 mL TOP-Se (2 M), and 0.375 mL ODE. Solution (B) was obtained by mixing 1.25 mL Cd-oleate, 1.25 mL TOP-Se, and 2.5 mL ODE. Solution (C) was obtained by mixing 1 mL of TOP-Se and 1mL of TOP-S. Solution (D) was 0.5 mL of TOP-S.

***Synthesis.*** The growth of CdSe/Cd$_x$Zn$_{1-x}$Se/ZnSe$_{0.5}$S$_{0.5}$/ZnS QDs was conducted using a multi-step procedure illustrated in Supplementary Fig. 1. To grow CdSe cores, a mixture of 6 mL ODE and 0.2 mL Cd-oleate was loaded into a three-neck flask and degassed at 120 °C for 15 min. The reaction was then saturated with nitrogen and heated to 310 °C. When the temperature reached 310 °C, 0.1 mL, TOP-Se was swiftly injected into the reaction flask. After 30 s, 1 mL TOP was added dropwise over 20 s. After 2 min, 1 mL of solution (A) was added dropwise to the reaction at the 5 mL/h rate over 12 min. At the end of the injection, CdSe CQDs with the band-edge (1S) absorption peak at 615 nm were obtained.

To start the growth of the compositionally graded Cd$_x$Zn$_{1-x}$Se layer, 2 mL of Zn-oleate solution were injected at once and 5 mL of solution (B) were added continuously over the course of 75 min at the 4 mL/h rate. During this stage, Zn-oleate solution was added shot-wise *via* three quick injections in the amounts 2, 4, and 2 mL at *ca*. 18.75, 52.5, and 67.5 min after the start of the graded-layer growth. These time intervals corresponded to 1.25, 3.5, and 4.5 mL of solution (B) fed into the reaction.

To grow the ZnSe$_{0.5}$S$_{0.5}$ layer, 2 mL of solution (C) was continuously injected with the 1 mL/h rate over 120 min. Zn(OA)$_2$ was added in several shots in the amount of 2 mL per shot for every 0.25 mL of (C) fed into the reaction (time intervals between consecutive injections were approximately 15 min). After addition of (C) was complete, solution (D) was continuously injected over the course of 30 min at the 1 mL/h rate. During this time, 1 mL Zn(OA)$_2$ was added shot-wise every 0.25 mL of (D) (that is, every 15 min). The last reaction step produced the final ZnS protective layer. To complete the reaction, the heating mantle was removed and the reaction products were cooled down to room temperature. This synthesis resulted in ccg-QDs with the following dimensions: 2.6 nm (core radius), 2.4 nm (graded Cd$_x$Zn$_{1-x}$Se layer thickness), 1 nm (ZnSe$_{0.5}$S$_{0.5}$ layer thickness), and 0.25 nm (final ZnS shell thickness).

***Purification***. Purification of the synthesized ccg-QDs QDs was carried out by diluting the content of the reaction flask with 35 mL chloroform and, then, adding 70 mL of ethanol as an antisolvent. This mixture was centrifuged at 6000 rpm for 10 min and, then, the precipitate was dissolved in 10 mL toluene. These solution-based ccg-QD samples were used in spectroscopic studies.

For fabrication of devices, QD samples were further purified using a multi-step procedure. Specifically, 4 mL of ccg-QDs in toluene were mixed with 30 mL acetone. The mixture was centrifuged at 7000 rpm for 15 min. The precipitate was dissolved in 2 mL hexane and, then, after addition of 20 mL of acetone the mixture was centrifuged at 7000 rpm for 15 min. The precipitate was dissolved again in 1 mL hexane. This was followed by addition of 15 mL of acetone and centrifugation at 7000 rpm for 15 min. The dry precipitate was weighted and the purified ccg-QD sample was dispersed in octane to obtain the desired concentration (typically, 50 mg mL$^{-1}$).

**Fabrication of LEDs**

Electrodes made of low index ITO (L-ITO) deposited onto a glass substrate were purchased from Thin Film Devices Inc. To fabricate an LED, the L-ITO electrode was cleaned *via* sequential 10-min sonication steps using isopropyl alcohol, acetone, and ethanol. After this procedure, the solvents were removed by 'baking' the electrode in a hot oven at 120 °C. Afterwards, 20 µL of ccg-QD solution (50 mg mL$^{-1}$) were spin-coated onto the L-ITO electrode at 2,000 rpm for 30 sec to form 2 monolayers of the ccg-QDs. This procedure was repeated twice to produce a 6-monolayer-thick ccg-QD film. During the deposition, the film was allowed to dry between consecutive spin-coating steps. Afterwards, a 50-nm HTL of Spiro-2NPB was deposited by thermal evaporation under vacuum (<10$^{-6}$ torr) using the evaporation rate of 0.3 Å s$^{-1}$. To prepare 'current-focusing' LEDs, a 60-nm thick LiF interlayer was thermally evaporated onto the Spiro-2NPB HTL using a shadow mask with the 50-µm gap. Then, a 10-nm-thick HIL of HAT-CN was deposited by thermal evaporation (0.2 Å s$^{-1}$ rate). The device was completed with a 100-nm-thick Ag electrode prepared by thermal evaporation (1 Å s$^{-1}$ rate) through a shadow mask with a 300-µm slit orthogonal to the opening in the LiF interlayer for additional 'current-focusing'.

**Characterization of LEDs**

The LEDs were characterized at both room and LN temperature. For LN measurements, the devices were loaded into a cryostat (Janis ST-100) adapted for electro-optical measurements. A function generator (Tektronix AFG 320) was used to produce square-shaped voltage pulses (0.1 - 3.5 V amplitude) at a desired repetition rate. The pulses were amplified by a high-speed bipolar amplifier (HSA4101, NF Corporation) with the 20X gain. The applied voltage was measured using a Tekronix (TDS 2024B) oscilloscope connected to the monitoring port of the amplifier. The current was monitored using the same oscilloscope by measuring a voltage drop across a 10 Ω load resistor on the current return. The device EL emitted through the L-ITO/glass electrode was measured using an imaging system, consisting of two lenses, a Czerny-Turner spectrograph (Acton SpectraPro 300i), and an LN-cooled charge-coupled device (CCD) camera (Roper Scientific).

**Spectroscopic measurements**

***Optical absorption measurements.*** Optical absorption spectra were measured using an integrating sphere module of a UV/Vis spectrometer (Lambda 950, Perkin Elmer). ITO and low index-ITO samples were measured as purchased. All HTL samples were prepared on bare glass substrates by thermal evaporation at the 0.2 – 0.3 Å s$^{-1}$ a rate. For calibration, a bare glass was measured and the acquired spectrum was used as a background signal.

***Optically excited ASE.*** Prior to ASE measurements, the devices were cleaved using a diamond tip to allow access to emission radiated from the edge of the active ccg-QD layer. Importantly, following the cleaving procedure, the devices remained EL active. A cleaved device was mounted into a cryostat and characterized at either room or LN temperature. Optical excitation was conducted *via* an opening in the sample holder through the transparent L-ITO/glass electrode. A pulsed regeneratively-amplified ytterbium-doped potassium gadolinium tungastate (Yb:KGW) laser (Pharos, Light Conversion) was used as an excitation source. The 190-fs, 343-nm pump pulses (1-kHz repetition rate) were obtained by tripling the fundamental laser output at 1030 nm. A 10-cm cylindrical $CaF_2$ lens was used to focus the beam onto the device into a strip with dimensions 40 µm by 1.7 mm. The pump beam was clipped with a razor blade to ensure that the area excited by the laser was fully within the injection area defined by the current focusing structure. The excited photoluminescence (PL) of the active area was collected from the device cleaved edge with a 10 cm lens and spectrally resolved using a Czerny-Turner spectrograph (Acton SpectraPro 300i) coupled to a LN-cooled CCD camera (Roper Scientific).

***VSL measurements.*** VSL measurements were conducted using the same configuration as in the ASE experiments (see above) with the excited stripe-shaped area of a varied length. The stripe length ($x$) was varied in ~20 µm steps using a razor blade mounted onto a translation stage. Optical gain ($G$) was obtained by fitting the measured edge-emitted PL intensity ($I$) to $I = A[\exp(Gx) – 1]/G + Bx$, where $A$ and $B$ were $x$-independent constants.

***PL lifetime measurements.*** A QD sample, prepared as either a spun-coated film or a solution loaded into a 1-mm thick quartz cuvette, was excited by 343-nm, <190-fs pulses (200-kHz repetition rate) focused into an 80-μm-diameter spot. The emitted PL was collected in the direction normal to the sample plane, spectrally dispersed with a Czerny-Turner spectrograph (Acton SpectraPro 300i) and detected with an avalanche photodiode (Micro Photon Devices) coupled to a time-correlated single-photon counting system (PicoQuant PicoHarp).

**Photonic modeling**

*Methodology.* To conduct photonic modeling of waveguided modes, a wave optics module of COMSOL Multiphysics software was used. Electromagnetic Waves, Wavelength Domain was selected for parametric sweep calculations with the wavelength ranging from 500 nm to 650 nm. The studied LED structures were considered as an ideal planar slab waveguide with zero roughness. All computation was conducted in time-independent two-dimensional domain.

*Multi-layered device geometry.* An LED structure was modeled by stacking rectangles whose thicknesses were defined by parameters of experimentally realized devices. To simulate an LED with a conventional inverted architecture, the following structure was considered: glass (1 μm)/ ITO (150 nm)/ZnO (50 nm)/QD (95 nm) / TCTA (40 nm)/MoO$_x$ (10 nm)/Ag (100 nm) (Fig. 2c). The novel architecture was implemented using the following arrangement of device layers: glass/L-ITO (250 nm)/QD (95 nm)/Spiro-2NPB (50 nm)/HAT-CN (10 nm)/Ag (100 nm) (Fig. 2d).

*Optical characteristics of device layers.* Complex refractive indices ($n,k$) of device layers were defined as a function of wavelength. Refractive indices of glass, ITO, ZnO, MoO$_3$, and Ag were extracted directly from the COMSOL material library. A refractive index of a QD solid was adapted from the previous study of cg-QD films.[1] The QD layer was assumed to be optically transparent ($k = 0$), which corresponded to the condition of the optical gain threshold. Based on the measured optical absorption spectra of organic HTLs (Fig. 2b), their extinction coefficient was also assumed to be zero. Refractive indices of organic HTLs were adapted from previously measurements of ref 38.

*Simulations.* Simulations were conducted by solving wavelength domain Maxwell's equations *via* a finite element method procedure applied to the selected device geometry:

$$\nabla \times (\mu_r^{-1} \nabla \times E) - k_0^2 \varepsilon_r E = 0$$

where $\mu_r$ is unity (no magnetic materials) and $\varepsilon_r$ (relative permittivity) was defined as $(n - jk)^2$.

***Mode confinement factors.*** A three-component electric-field vector (**E**) was obtained in the 2D geometric domain as a result of computation. A mode confinement factor of layer *i* was calculated from

$$\Gamma_i = \frac{\int_i |\mathbf{E}|^2 \mathrm{d}x\mathrm{d}y}{\int |\mathbf{E}|^2 \mathrm{d}x\mathrm{d}y}$$

where the integration volumes extended over the selected layer (numerator) and the entire space (denominator).

***Optical loss coefficients.*** Two approaches were used to obtain total loss coefficients. In one approach, the effective refractive index ($n_{\text{eff}}$), obtained from the boundary mode analysis, was related to the total loss coefficient ($\alpha_{\text{loss}}$) of a multi-layered stack by

$$\alpha_{\text{loss}} = -\frac{4\pi}{\lambda} \times \mathrm{Im}(n_{\text{eff}})$$

In the second approach, the waveguide transmittance (*T*) was obtained based on the signal at the output port. Then, using the waveguide length (*L*), the total loss coefficients was calculated based on the Beer-Lambert law from

$$\alpha_{\text{loss}} = -\ln(T)/L$$

The results obtained by these two methods were different by less than 0.2%.

***Spatial profile of the electromagnetic power density loss.*** A cross-sectional profile of the power-density loss (optical loss) within a waveguide was calculated from

$$Q_{\mathrm{e}} = \frac{1}{2}\omega\varepsilon_0\varepsilon_{\mathrm{r}}''|\mathbf{E}|^2$$

where $\omega$, $\varepsilon_0$, and $\varepsilon_{\mathrm{r}}''$ were the angular frequency of the propagating wave, the vacuum permeability, and the imaginary part of the relative permittivity, respectively.

**Data availability**

The data that support the findings of this study are available on request from the corresponding author V.I.K.

## Acknowledgements

This project was supported by the Laboratory Directed Research and Development (LDRD) program at Los Alamos National Laboratory under projects 20200213DR and 20210176ER. N.A. acknowledges support by a LANL Director's Postdoctoral Fellowship.

## Author Contributions

V.I.K. initiated the study, coordinate the project, and wrote the manuscript together with N.A., and C.L. N.A designed, fabricated, and characterized the low-loss, high-gain electroluminescent devices. Y.-S.P. and C.L. performed spectroscopic studies of the ccg-QDs including ASE measurements. Y.-S.P. measured modal gain coefficients using a VSL technique. J.D. synthesized the ccg-QDs. N.A. conducted COMSOL photonic modeling of multilayered devices and performed their optimization. N.A. and Y.-S.P. contributed equally to this work.

## Competing financial interests

The authors declare no competing financial interests.

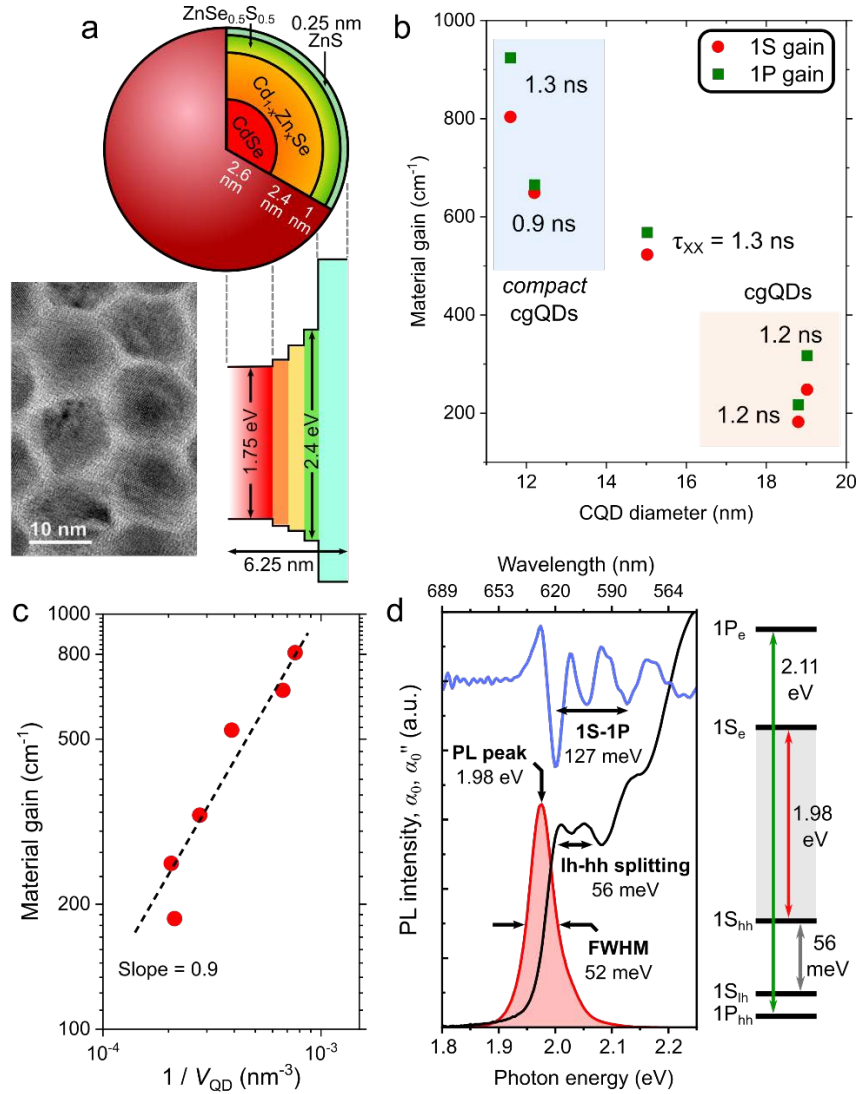

**Figure 1. Structural, electronic, and optical characteristics of ccg-QDs. a,** The internal structure of the CdSe/Cd$_{1-x}$Zn$_x$Se/ZnSe$_{0.5}$S$_{0.5}$/ZnS ccg-QD (2.6-nm core radius, 6.25 nm overall radius) (top), their exemplary transmission electron microscopy (TEM) image (bottom left), and the approximate shapes of the 'graded' electron and hole confinement potentials (bottom right). **b,** The 1S (red circles) and 1P (green squares) material gain coefficients of the cg-QDs as a function of the overall diameter. Biexciton lifetimes ($\tau_{XX}$) are indicated next to corresponding data points. **c,** The measured 1S material gain (red circles) scales approximately linear with inverse of the overall QD volume (dashed black line). **d,** (Left) The spectra of linear absorption (black line), its second derivative (blue line), and PL (red line) of the ccg-QD sample displayed in '**a**'. (Right) The structure of the near-band-edge electron (1S$_e$, 1P$_e$) and hole (1S$_{hh}$, 1S$_{lh}$, 1P$_{hh}$) states along with approximate inter-state energies derived from spectra on left. All experimental data shown in this figure were obtained at room temperature.

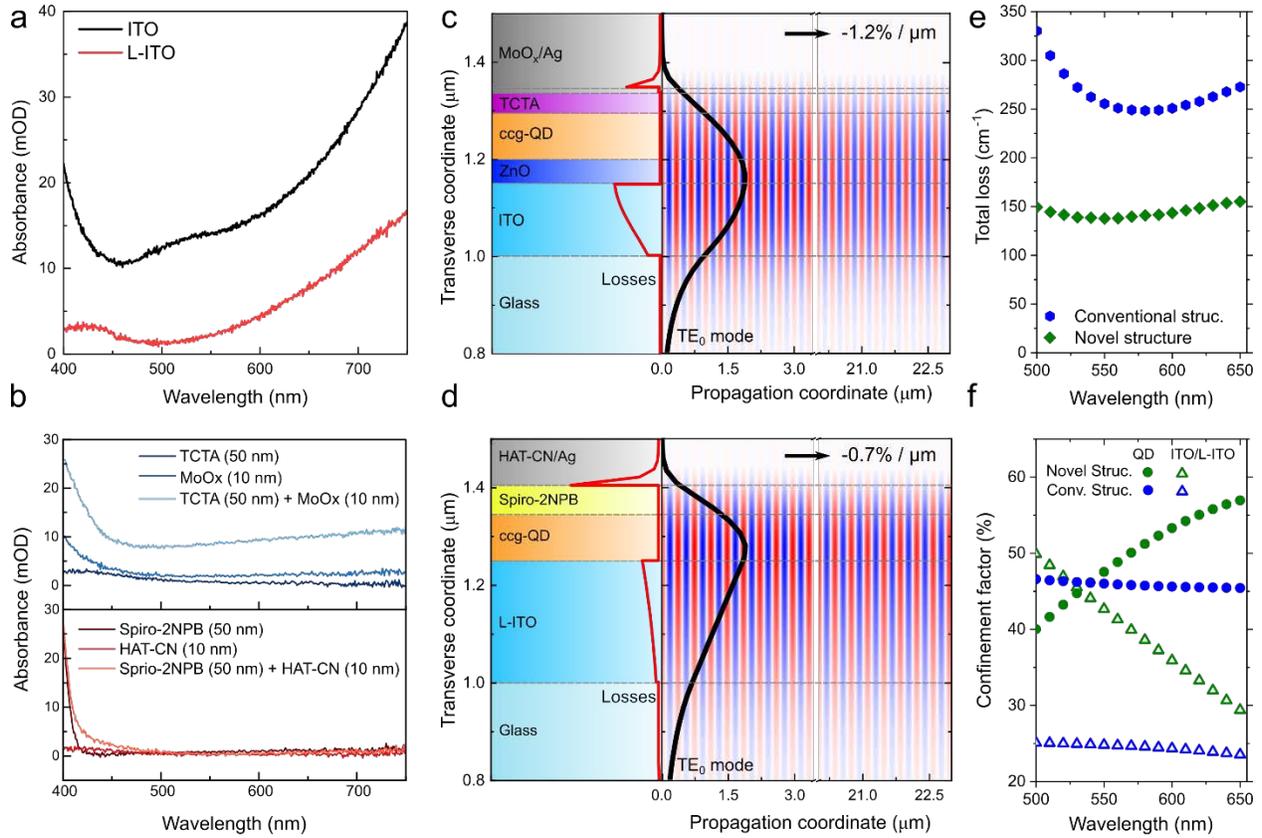

**Figure 2. Photonic properties of the conventional and the novel low-loss LED architecture. a,** Optical absorption spectra of thin films of standard ITO (black) and L-ITO (red) of identical thicknesses of ~150 nm prepared on glass substrates. **b,** Optical absorption spectra of HTLs and HILs used in conventional LEDs (top) in comparison to those employed in the novel devices (bottom). The data are shown for both the individual HTLs (50 nm) and HILs (10 nm) as well as the HTL/HIL bilayers (50 nm + 10 nm). **c,** COMSOL simulations of optical losses and $TE_0$ mode propagation along the active ccg-QD layer in a conventional LED (glass/ITO (150 nm)/ZnO (50 nm)/ ccg-QDs (95 nm)/TCTA (40 nm)/$MoO_x$ (10 nm)/Ag (100 nm). **d,** Same for the novel device architecture (glass /L-ITO (250 nm) / ccg-QD (95 nm) /Spiro-2NPB (50 nm)/HAT-CN (10 nm)/Ag (100 nm). **e,** Simulated loss coefficients for the conventional (blue circles) and the novel (green diamonds) LED architecture. **f,** Simulated confinement factors of the optical $TE_0$ mode for the ccg-QD (circles) and the ITO (triangles) layers in the case of the conventional (blue symbols) and the novel (green symbols) LED architecture.
27

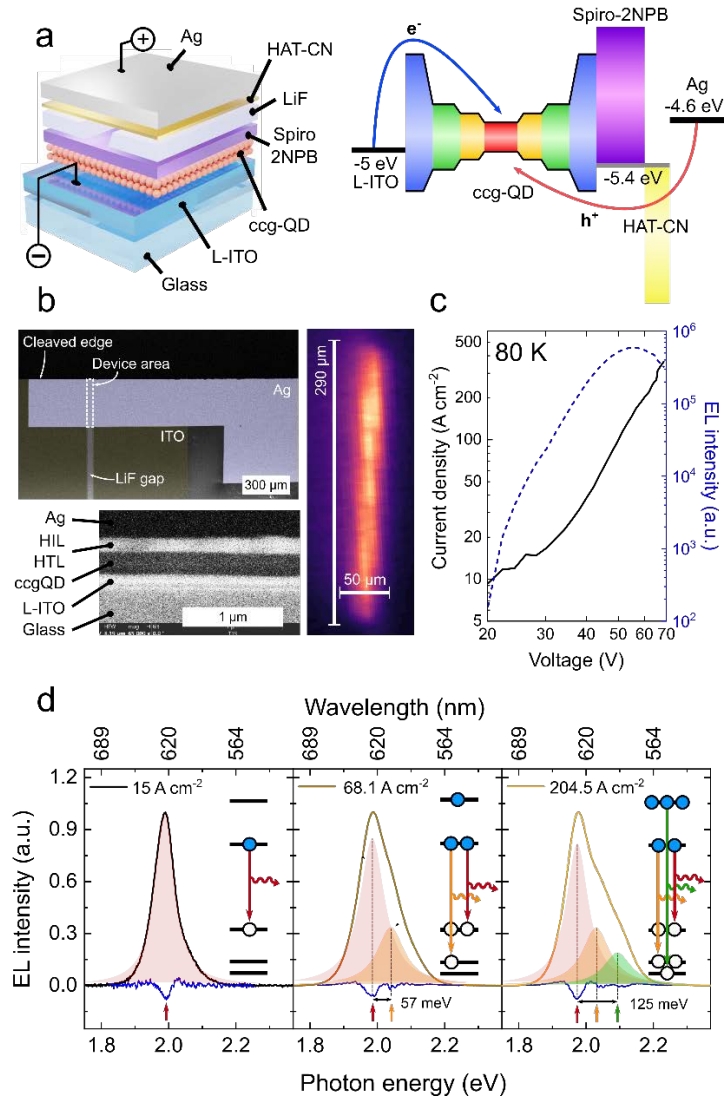

**Figure 3. Practical implementation of low-loss LEDs and their EL properties. a,** A schematic depiction of the low-loss LED device stack analyzed in Fig. 2d (left) and the corresponding band diagram (right). **b,** Top- and side-view SEM images of the fabricated device (top left and bottom left, respectively) wherein the injection area (highlighted by the white line) is defined by the intersection of the Ag electrode and the gap in the LiF interlayer. As a result of this 'current-focusing' design, the emitting area is confined to *ca.* 290 μm by 50 μm (image at right). **c,** The *j-vs.-V* (black line, left axis) and EL intensity-*vs.-V* (blue line, right axis) characteristics of the fabricated LED measured at LN temperature. The device is driven using electrical pulses with the 500-ns duration and the 500-Hz repetition rate. **d,** Normalized EL spectra measured at LN temperature (black lines) and the spectra of their 2$^{nd}$ derivative (blue lines) for current densities of 15 (left), 68.1 (middle) and 204.5 (right) A cm$^{-2}$. These spectra can be accurately described using 1 to 3 Lorentzian bands (coloured shading) that correspond to the $1S_e-1S_{hh}$ (red arrows), $1S_e-1S_{lh}$ (orange arrows), and $1P_e-1P_{hh}$ (green arrows) transitions (insets). These bands emerge one after the other when the QD is excited with more than 0, 2, and 4 e-h pairs, respectively (electrons and holes are shown in the insets by solid and open circles, respectively).

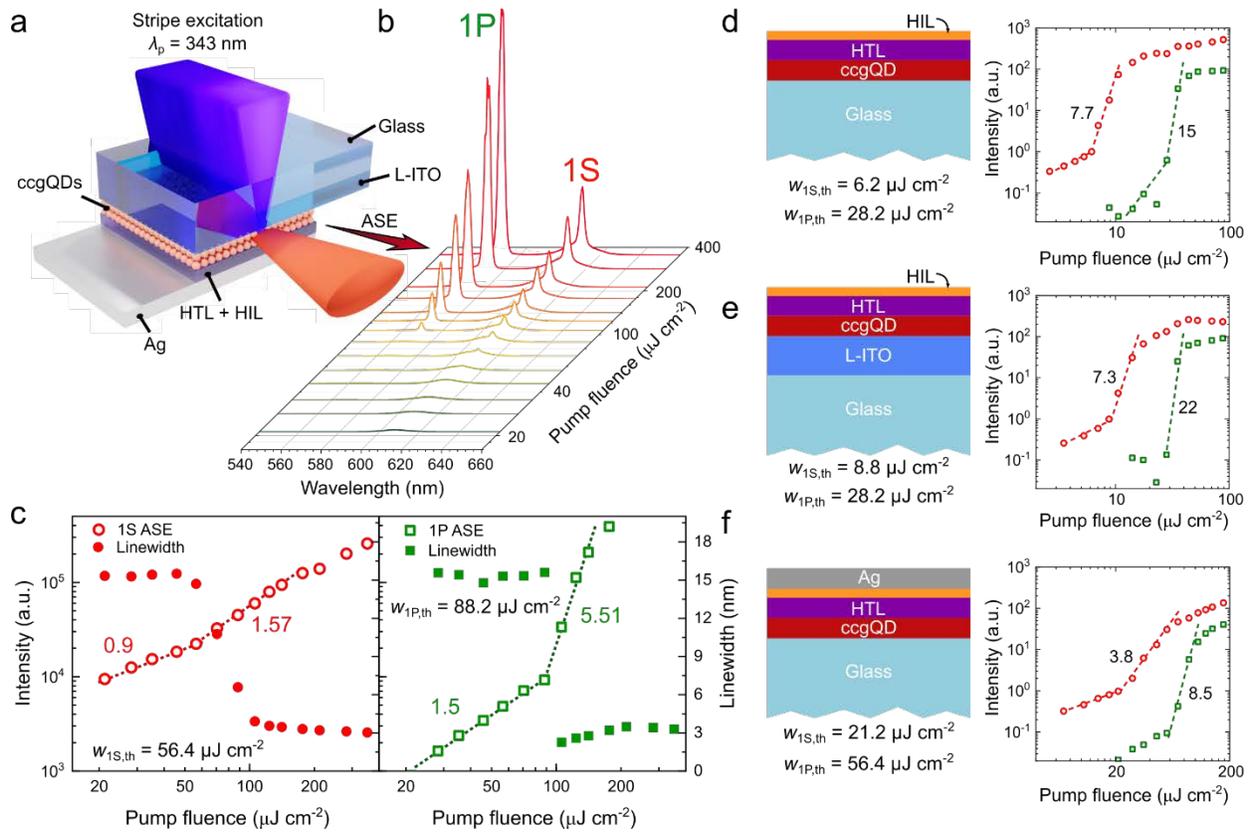

**Figure 4. Realization of two-band ASE in a fully stacked LED. a,** A configuration used in the ASE measurements. The ccg-QD layer is excited with 343-nm femtosecond laser pulses through the L-ITO device side and the resulting PL is collected from the device edge. The optically excited area is shaped so as to fit within the charge-injection area. **b,** The PL spectra recorded as a function of pump fluence exhibit ASE due to both the 1S and the 1P transitions. **c,** The 1S (left, red open circles) and 1P (right, green open squares) PL intensities and linewidths (solid symbols) for the increasing pump fluence. These data exhibit typical signatures of the transition to ASE signaled by the linewidth narrowing and the change in the log-log slope of the growth of the PL intensity. Based on these measurements, ASE thresholds are 56.4 (1S) and 88.2 (1P) μJ cm$^{-2}$. **d-f,** The schemes of three partial device stacks (left) and the corresponding 1S (red circles) and 1P (green circles) PL intensities measured as a function of pump fluence. The transition to ASE is evident from a sharp change in the log-log slope of the measured dependences. Post-ASE-threshold slopes are indicated in the figure. All measurements were performed at LN temperature.

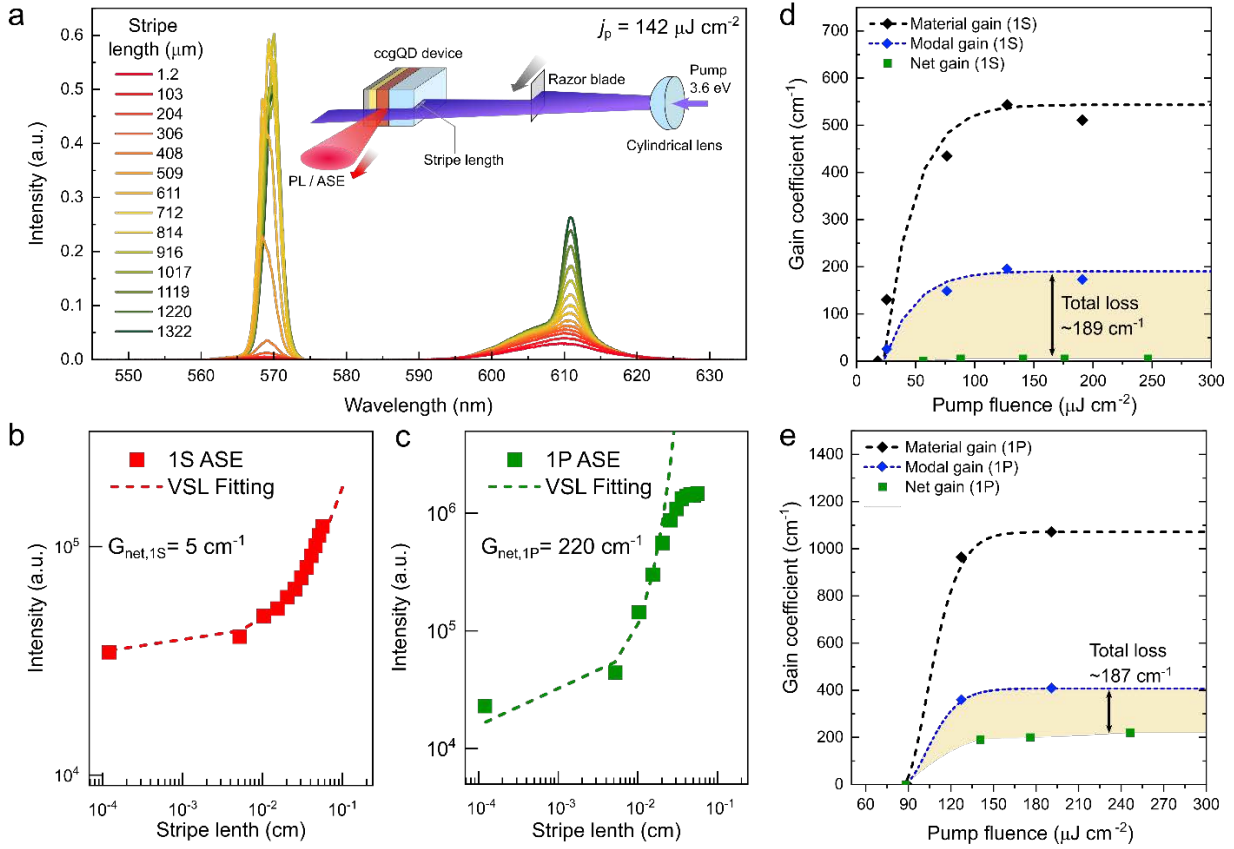

**Figure 5. Measurements of the optical gain and loss coefficients for the complete EL-active device. a,** Device-based VSL measurements (inset) display pronounced signatures of ASE for both the 1S and the 1P emission bands. The ccg-QD layer is excited using 3.6-eV femtosecond laser pulses with the pump fluence of 142 μJ cm$^{-2}$. **b&c,** The 1S (**b**) and 1P (**c**) PL intensities (symbols) as a function of stripe length measured for $w_p$ = 248 μJ cm$^{-2}$. Based on the standard VSL analysis (dashed lines), the net 1S and 1P gain coefficients are 5 and 220 cm$^{-1}$, respectively. **d,** The 1S material, modal, and net gain coefficients (black diamonds, blue diamonds, and green squares, respectively) obtained from the pump-fluence-dependent VSL measurements of the thick ccg-QD film on a glass substrate and the complete device; dashed lines are guides for the eye. The difference between the modal and the net gain coefficients (yellow shading) yields the total loss coefficient of 185 cm$^{-1}$. **e,** A similar set of data for the 1P transition yields $\alpha_{loss}$ of 187 cm$^{-1}$. All measurements were performed at LN temperature.

**Table 1.** Optical gain and loss coefficients for the complete EL-active device derived from the LN VSL measurements in the regime of gain saturation (Fig. 5).

| Coefficient (cm$^{-1}$) | Material gain ($G_{mat}$) | Modal gain ($G_{mod} = G_{mat}\Gamma_{QD}$) | Total loss ($\alpha_{loss}$) | Net gain ($G_{net} = G_{mod} - \alpha_{loss}$) |
|---|---|---|---|---|
| 1S transition | 543 | 190 | 185 | 5 |
| 1P transition | 1071 | 407 | 187 | 220 |

**Supplementary Information for**

**Optically Excited Two-Band Amplified Spontaneous Emission from a High-Current-Density Quantum-Dot LED**

Namyoung Ahn[1†], Young-Shin Park[1,2†], Clément Livache[1], Jun Du[1,3], and Victor I. Klimov[1]*

[1]Nanotechnology and Advanced Spectroscopy Team, C-PCS, Chemistry Division, Los Alamos National Laboratory, Los Alamos, New Mexico 87545, USA

[2]Department of Chemical and Biomolecular Engineering, Korea Advanced Institute of Science and Technology, Daejeon 34141, Republic of Korea

[3]State Key Laboratory of Molecular Reaction Dynamics, Dalian Institute of Chemical Physics, Chinese Academy of Sciences, Dalian, Liaoning 116023, China

*klimov@lanl.gov

**Supplementary Note 1. Ideal optical gain thresholds**

To evaluate optical gain thresholds for an active medium based on compact continuously graded quantum dots (ccg-QDs), we take into consideration 2 lower-energy electron states ($1S_e$ and $1P_e$) and 3 lower-energy hole states ($1S_{hh}$, $1S_{lh}$, and $1P_{hh}$) depicted in Fig. 1d of the main article. We further assume the situation of zero temperature ($T = 0$) when thermal excitation of charge carriers to higher energy states is absent. The optical gain threshold (that is, the onset of population inversion) for the transition, which couples the $i$-electron and the $j$-hole states, is defined by

$$f_{e,i} + f_{h,j} = 1,$$

where $f_{e,i}$ and $f_{h,j}$ are the occupation factors of each of the corresponding degenerate levels. The degeneracy factors of the S and P levels are $g_S = 2$ and $g_P = 6$, respectively.

Next, we assume that the QDs are charge neutral, that is, the QD electron occupancy is equal to the hole occupancy: $N_e = N_h = N$. If $N \leq 2$, the $1S_e$ and $1S_{hh}$ occupation factors are $f_{e,1S} = f_{h,1S} = N/g_S = N/2$. Hence, the optical gain threshold for the band-edge $1S_{hh}-1S_e$ transition ($N_{g,1Shh}$) can be found from

$$f_{e,1S} + f_{h,1Shh} = \frac{N_e}{2} + \frac{N_h}{2} = N = 1,$$

which yields $N_{g,1Shh} = 1$.

By applying the above approach to the $1S_e-1S_{lh}$ and $1P_e-1P_{hh}$ transitions, we obtain

$$f_{e,1S} + f_{h,1Slh} = \frac{N_e}{2} + \frac{N_h - 2}{2} = N - 1 = 1,$$

$$f_{e,1P} + f_{h,1Phh} = \frac{N_e - 2}{6} + \frac{N_h - 4}{6} = \frac{N}{3} - 1 = 1.$$

These expressions yield $N_{g,1Slh} = 2$ and $N_{g,1Phh} = 6$.

.

## Supplementary Figures

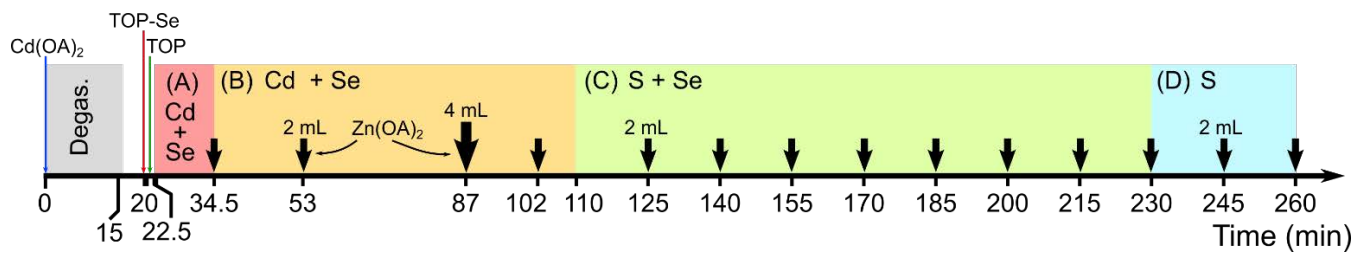

**Supplementary Figure 1. Schematic depiction of the multi-step ccg-QD synthesis and the corresponding timeline.**

The developed synthesis combines continuous injection of precursors (highlighted by colored shadings) with rapid (shot-wise) injections of $Zn(OA)_2$ (shown by vertical arrows). The red, yellow, green, and blue areas correspond to the growth of the CdSe core, the $Cd_{1-x}Zn_xSe$ graded layer, the $ZnSe_{0.5}S_{0.5}$ layer, and the final ZnS shell, respectively.

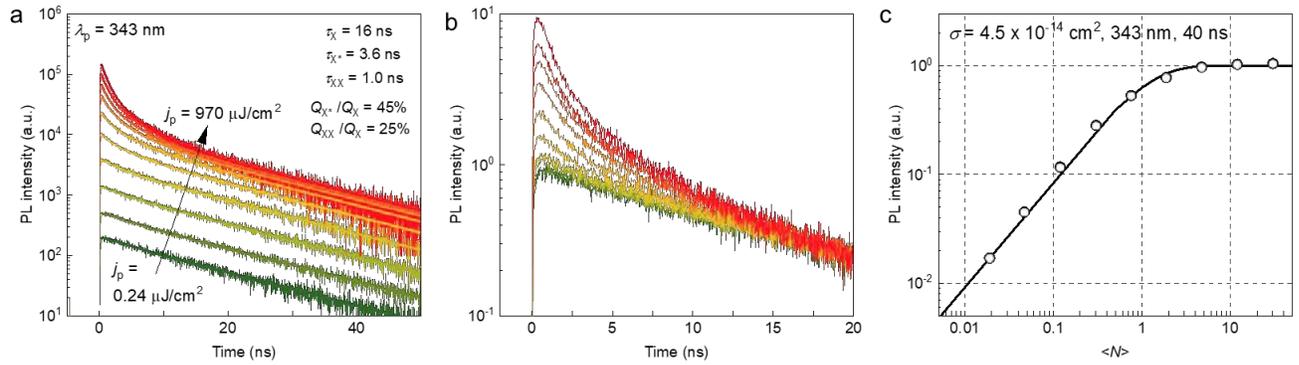

**Supplementary Figure 2. Photoluminescence (PL) dynamics of ccg-QDs.**

**a,** PL dynamics of the ccg-QD in toluene solution under pulsed excitation with the increasing pump fluence $w_p$ = 0.24 to 970 µJ cm$^{-2}$; the pulse duration is <190 fs and pump wavelength is 343 nm. Based on the global three-exponential fits of PL decays (smooth lines), the neutral-exciton, biexciton, and charged exciton (trion) lifetimes are $\tau_X$ = 16 ns, $\tau_{XX}$ = 1 ns, and $\tau_{X^*}$ = 3.6 ns. The biexciton and trion emission quantum yields ($Q_{XX}$ and $Q_{X^*}$), derived from $Q_{XX} = 4\tau_{XX}/\tau_X$ and $Q_{X^*} = 2\tau_{X^*}/\tau_X$ are 25% and 45%, respectively. These fairly high emissivities of the biexciton and trion states indicate strong suppression of Auger recombination in the ccg-QDs. **b,** Same PL dynamics as in **a** but presented in a 'tail-normalized' form to emphasize the long biexciton lifetime. **c,** The measured long-time PL intensity (time delay $t$ = 40 ns) as a function of pump fluence expressed in terms of the average QD excitonic occupancy, $\langle N \rangle$. Based on the Poisson fit (line) of the measured data (circles), the ccg-QDs absorption cross-section at 343 nm is $\sigma$ = 4.5·10$^{-14}$ cm$^2$. The fitting procedure employed the following expression: $I_{PL} = A [1 - \exp(-\sigma \langle N \rangle)]$, where $A$ is the constant, and $I_{PL}$ is the PL intensity.

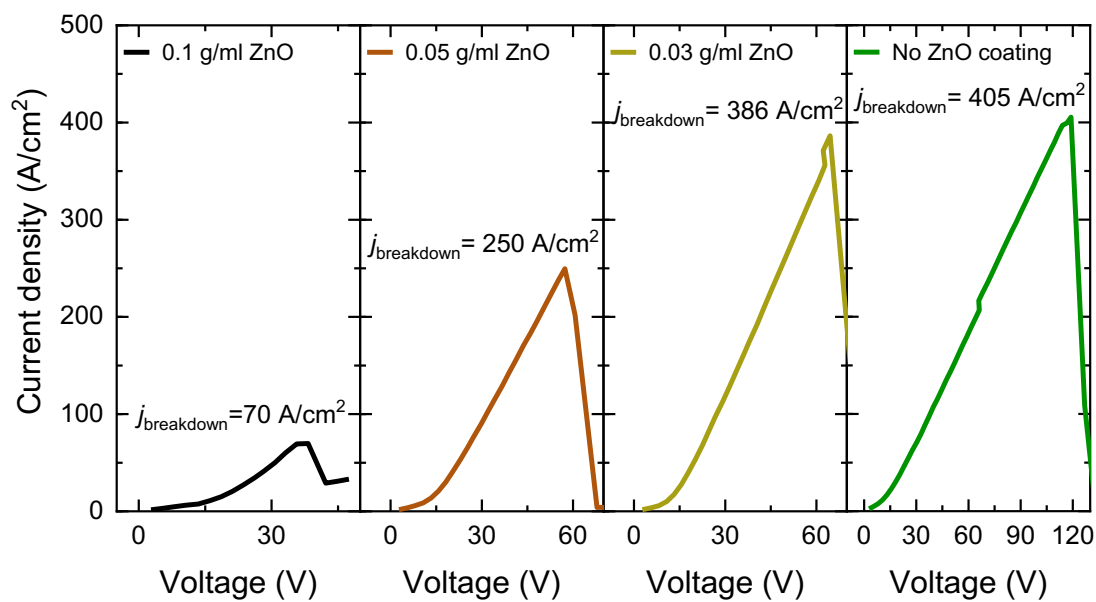

**Supplementary Figure 3. Effect of the ZnO layer on the breakdown current density of the ccg-QD based LED.**

The structure of the tested LEDs is L-ITO/ZnO/ccg-QDs/HTL/Ag. The ZnO layer of a varied thickness is prepared by spin-coating a ZnO precursor solution followed by annealing at 200 °C. The LEDs employing a thicker ZnO layer experience breakdown at lower current densities. The LED without the ZnO layer exhibit the highest breakdown current density of 405 A cm$^{-2}$. Due to its low thermal conductivity, the ZnO layer likely inhibits heat exchange with the substrate which increases device overheating and accelerates its degradation.

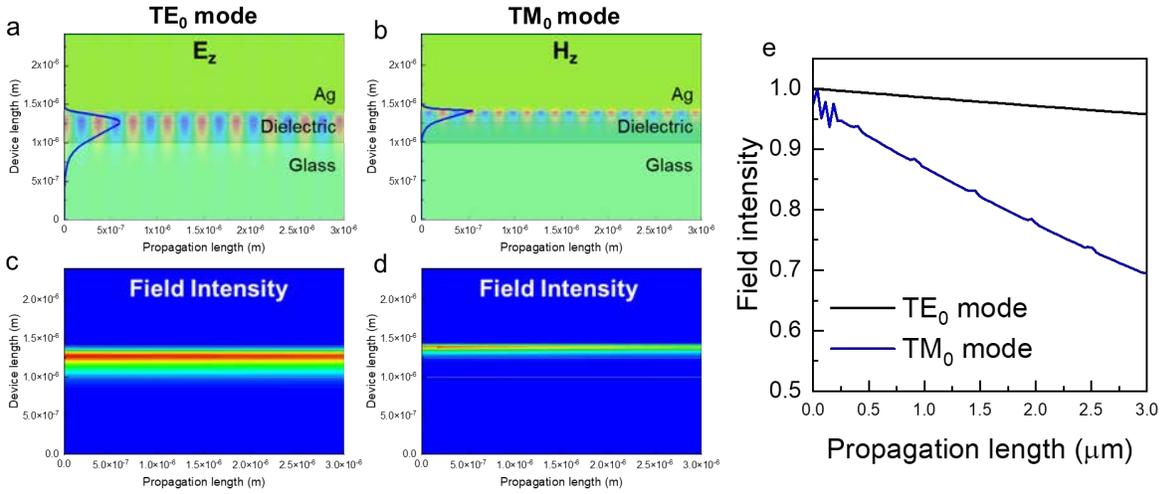

**Supplementary Figure 4. COMSOL simulations of transverse electric (TE) and transverse magnetic (TM) modes for a glass/Ag bilayer.**

**a,** A spatial distribution of the TE field ($E_z$) of the $TE_0$ wave ($\lambda$ = 600 nm) propagating along a dielectric/metal interface in a bilayer glass/Ag structure. The metal layer 'pushes' the optical field, leading to a photonic mode propagating primarily through a dielectric layer. **b,** A spatial distribution of the TM field ($H_z$) of the $TM_0$ wave for the same bilayer. A wave propagates along the metal surface as a plasmonic mode. **c&d,** Spatial distributions of field intensities for the $TE_0$ (c) and $TM_0$ (d) waves, respectively. **e,** The dependence of the $TE_0$ and $TM_0$ field intensities on propagation distance indicates a considerably faster attenuation of the $TM_0$ mode.

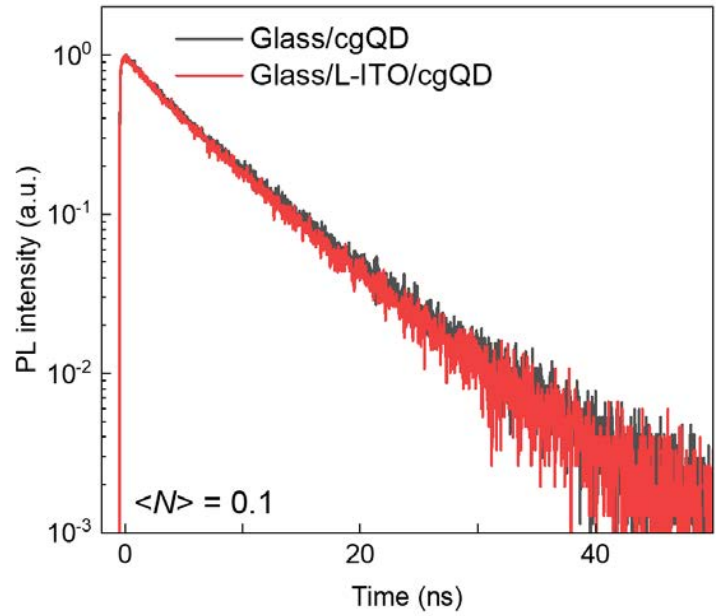

**Supplementary Figure 5. Effect of the L-ITO layer on ccg-QD emission dynamics.**

PL dynamics of the ccg-QDs films on the glass (black line) and the glass/L-ITO (red line) substrate obtained using low-intensity pulsed excitation at 3.6 eV. A close similarity between the two dynamics indicates that L-ITO does not quench ccg-QD emission.

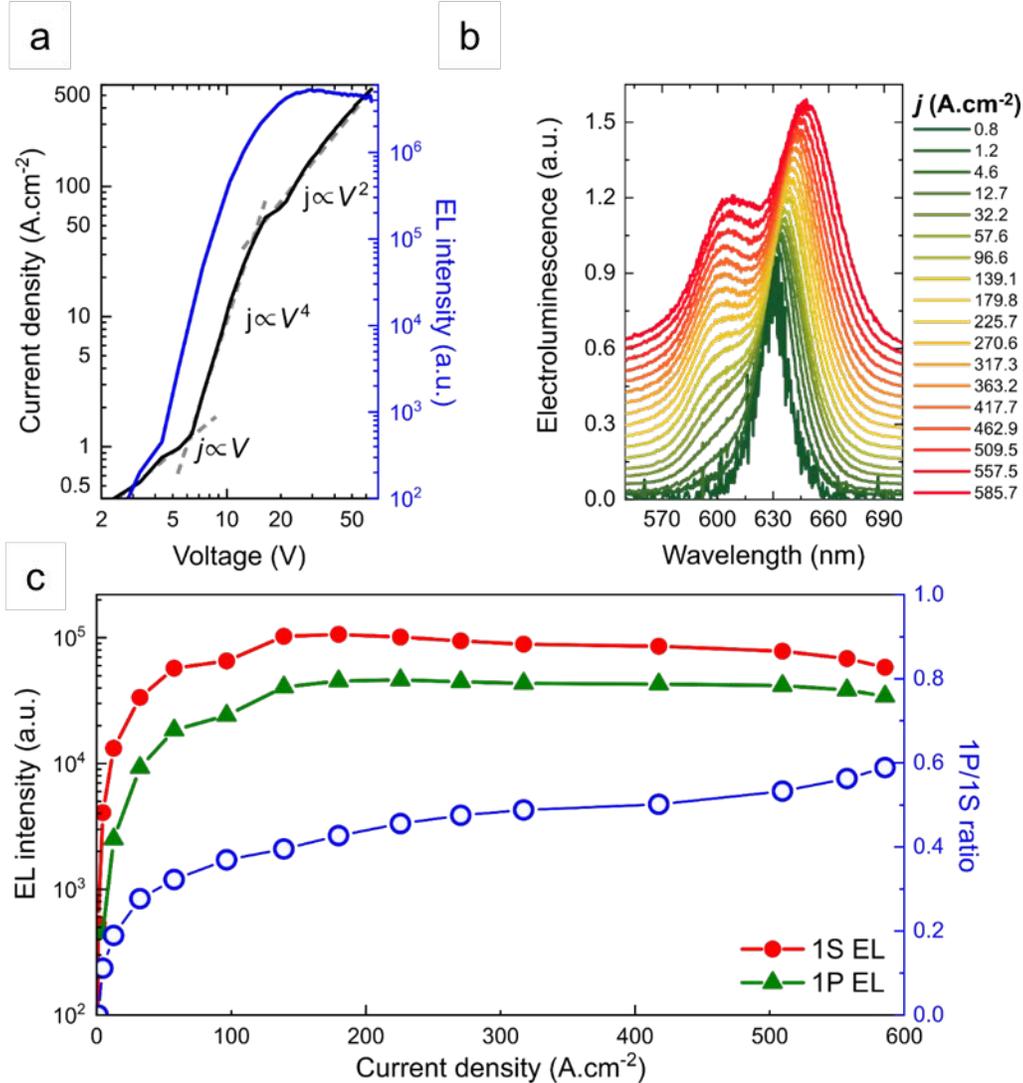

**Supplementary Figure 6. Room-temperature characteristics of the ccg-QD LED.**

**a,** The dependence of current density ($j$) versus bias ($V$) (black line, left axis) and the $V$-dependent electroluminescence (EL) intensity (blue line, right axis). The LED is driven by 5-µs voltage pulses at the 1-kHz repetition rate. The $j-V$ characteristics exhibits three different charge-transport regimes that can described by power dependence $j \propto V^m$ with $m$ = 1, 4, and 2 (see main article for details). **b,** The EL spectra plotted as a function of $j$ feature two emission bands. One is due to the band-edge 1S transition and the other (emerges at higher currents) is due to the higher-energy 1P transition. **c,** EL intensities for the 1S (red circles) and the 1P (green triangles) transitions. The 1P/1S intensity ratio ($r$) (blue circles, right axis) can be used to evaluate the number of the injected electron-hole pairs from $r = (N-2)^2/12$. Based on this expression, $N$ reaches ~4.7 at ~500 A cm$^{-2}$.

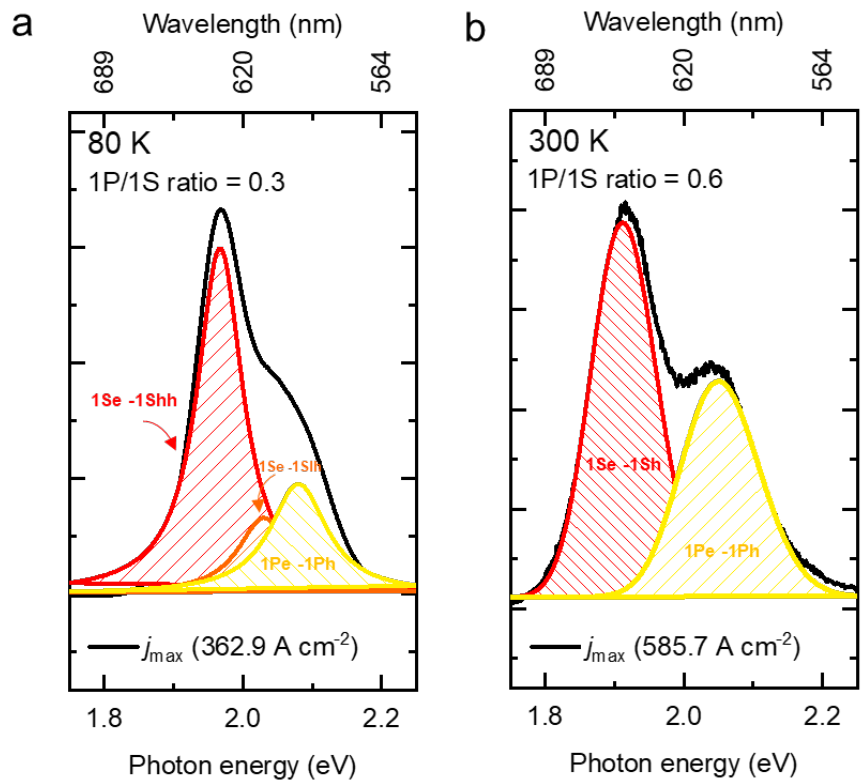

**Supplementary Figure 7. EL spectra at the maximal current injection before breakdown.**

**a,** The EL spectrum ($j$ = 362.9 A cm$^{-2}$) of the device operated at 80 K (black line) can be presented as a sum of three bands associated with three different QD transitions (indicated in the figure). Based on the band amplitudes, the 1P/1S intensity ratio is 0.3. **b,** The spectrum of room temperature EL ($j$ = 585.7 A cm$^{-2}$) shows the 1P/1S intensity ratio of 0.6.

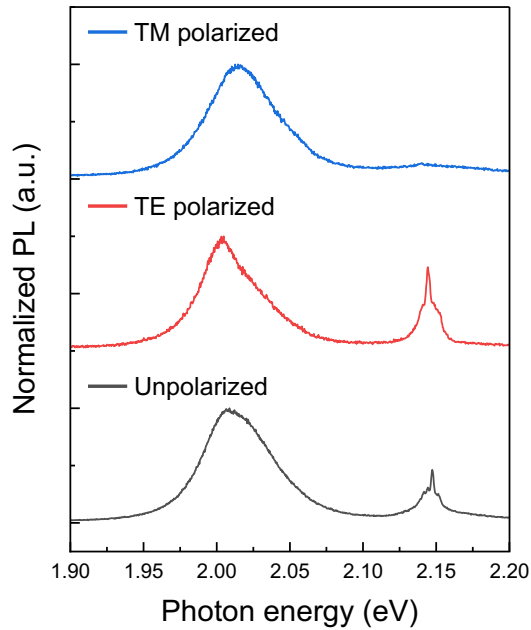

**Supplementary Figure 8. Polarization of amplified spontaneous emission (ASE) in a metal-covered waveguide.**

The spectra of emission from a cleaved edge of the glass/L-ITO/ccg-QD/HTL/HIL/Ag device under stripe-shaped optical excitation ($w_p$ = 200 µJ cm$^{-2}$) recorded without a polarizer (black), and with a linear polarizer aligned with either the TE polarization (red) or the TM polarization (blue). Both 1S and 1P ASE are TE-polarized, indicating that no ASE is supported by the TM waveguided mode due to strong propagation losses (see the results of COMSOL modeling in Supplementary Fig. 4).

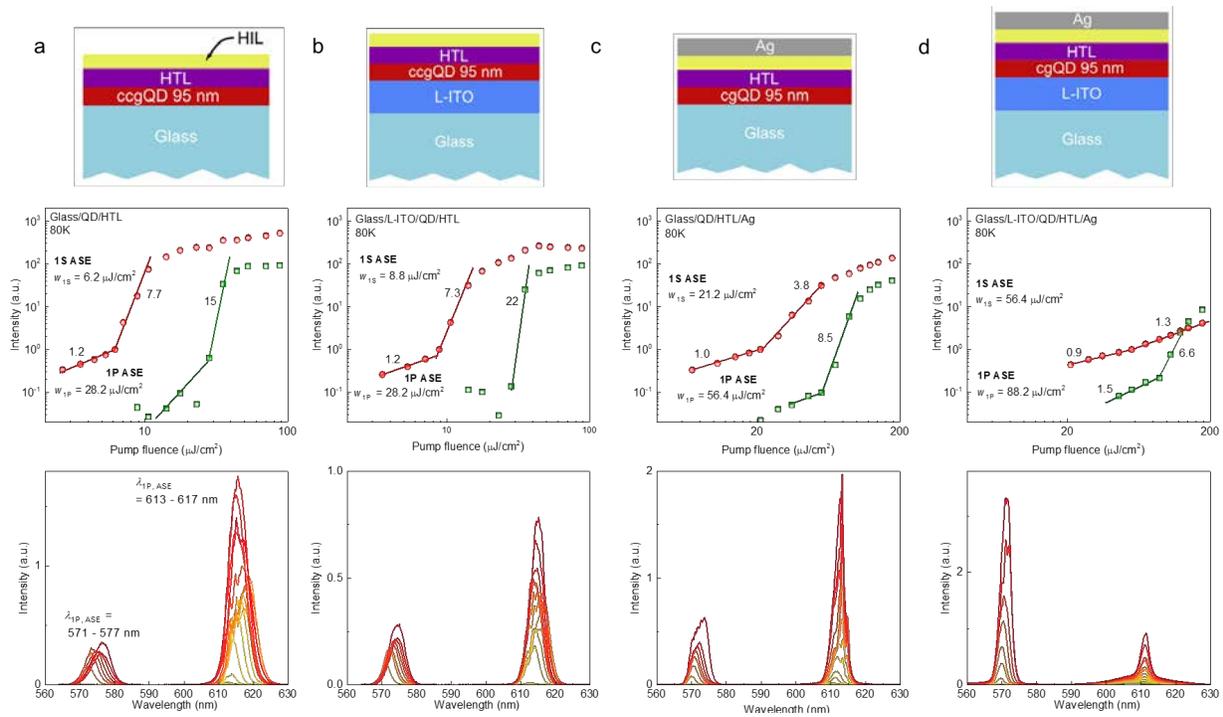

**Supplementary Figure 9. Optically excited ASE in partial device stacks.**

**a,** The ASE studies of the device stack comprising glass/ccg-QD/HTL/HIL (top). The 1S (red) and 1P (green) emission intensities (middle) derived from the pump-intensity-dependent spectra of edge-emitted light (bottom). **b,** Same for the glass/L-ITO/ccg-QD/HTL/HIL device. **c,** Same for the glass/ccg-QD/HTL/HIL/Ag device. **d,** Same for the glass/L-ITO/ccg-QD/HTL/HIL/Ag device. All measurements were conducted at $T$ = 80 K.

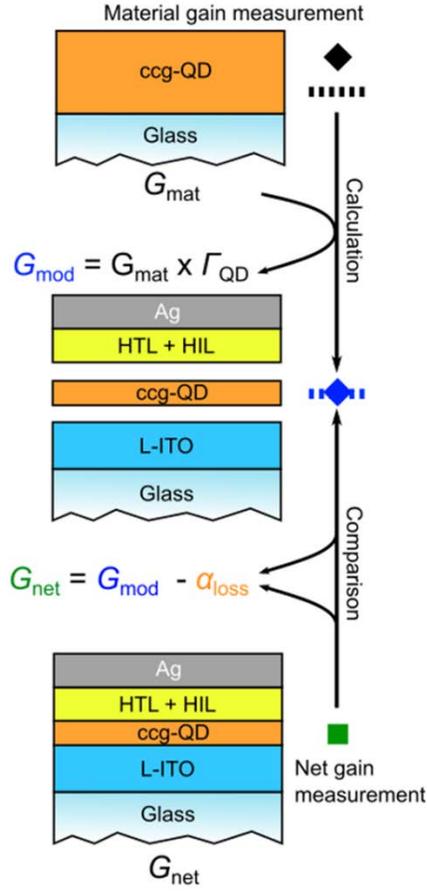

**Supplementary Figure 10. Evaluation of optical losses using variable stripe length (VSL) measurements.**

The diagram detailing the optical gain/loss quantification presented in Fig. 5 of the main article. Top: the material gain ($G_{mat}$) is inferred from the VSL measurements of a 300-nm thick layer of ccg-QDs assembled on a glass substrate. Middle: Modal gain ($G_{mod}$) is calculated as $G_{mod} = G_{mat}\Gamma_{QD}$, where $\Gamma_{QD}$ is the simulated confinement factor of the ccg-QD layer (Fig. 2d of the main article). Bottom: The net gain ($G_{net}$) is inferred from the VSL measurements of the complete EL-active device. The loss coefficient ($\alpha_{loss}$) is found from $\alpha_{loss} = G_{mod} - G_{net}$. Symbols on the right are color matched to symbols/lines in Fig. 5a,b of the main article.

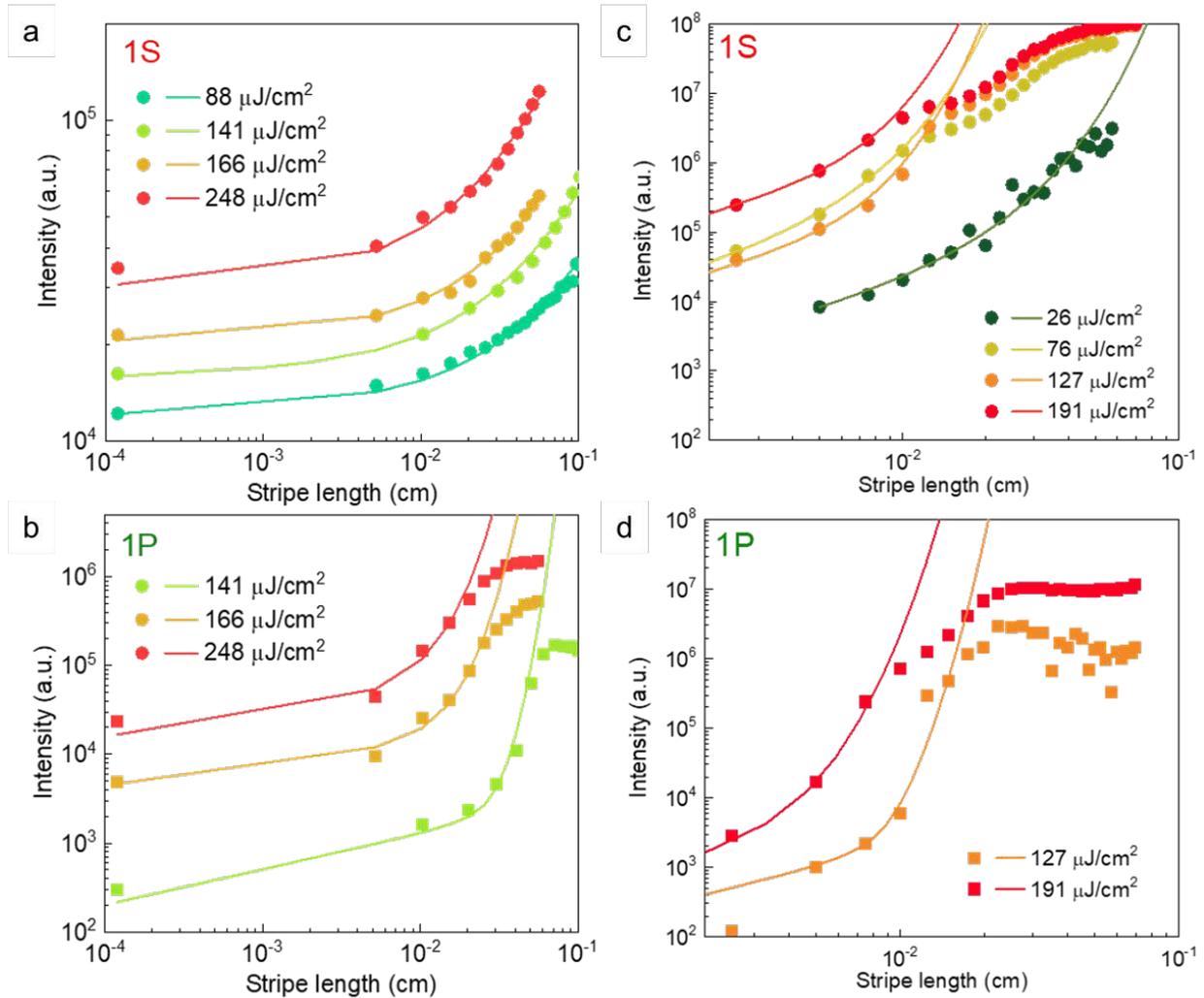

**Supplementary Figure 11. VSL measurements of ccg-QD films on the glass substrate and in the complete EL-active device.**

**a,** Pump-intensity-dependent VSL measurements of optical gain ($G$) of a 300-nm thick ccg-QD film on the glass substrate at the peak of the 1S emission band (circles). Lines are fits using $I = A(e^{Gx} - 1)/G + Bx$, where $x$ is the stripe length. **b,** Same for the 1P band. **c,** Same as in '**a**' but for the ccg-QD film in the complete EL-active device. **d,** Same as in '**c**' but for the 1P band. All measurements were conducted at $T = 80$ K.

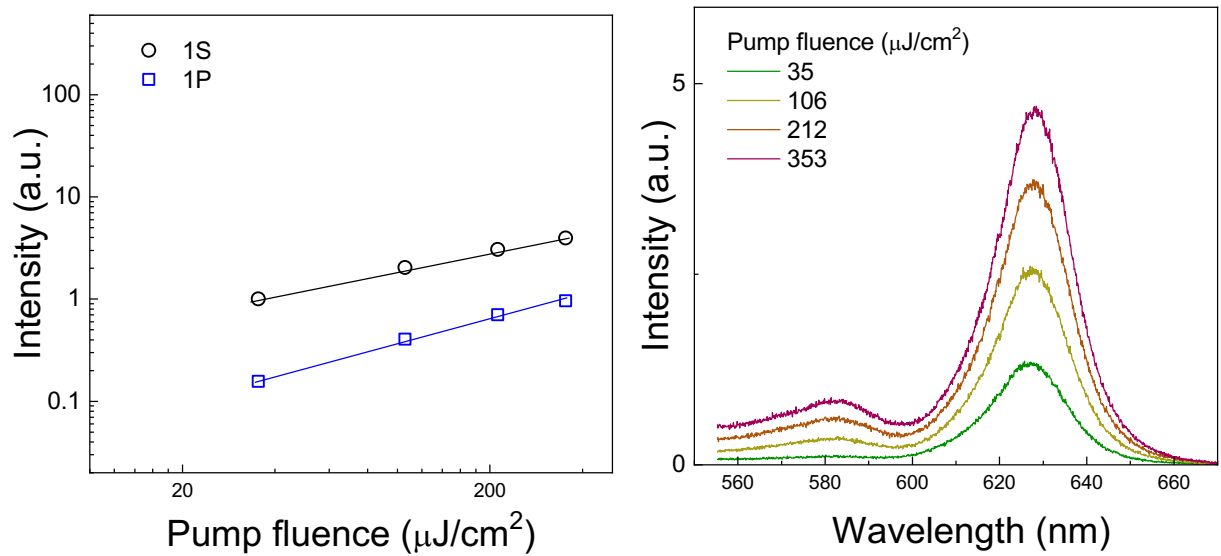

**Supplementary Figure 12. Room temperature VSL measurements of the complete EL device.**

The 1S (black) and 1P (blue) PL intensities (left) and the PL spectra (right) as a function of pump fluences. These measurements do not display any ASE, which is likely due to the temperature-induced increase in optical losses (see main article).

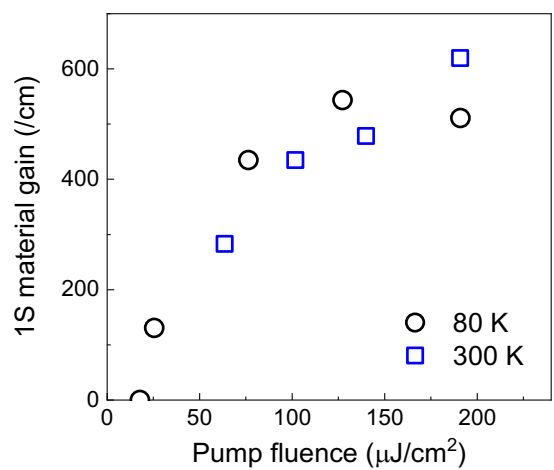

**Supplementary Figure 13. 1S material gain coefficients of ccg-QDs.**

The 1S material gain coefficient as a function of pump fluence measured using the VSL technique at 80 K (black circles) and 300 K (blue squares). Due to a wide separation between the band-edge and the higher-lying states, $G_{mat}$ is virtually temperature independent.

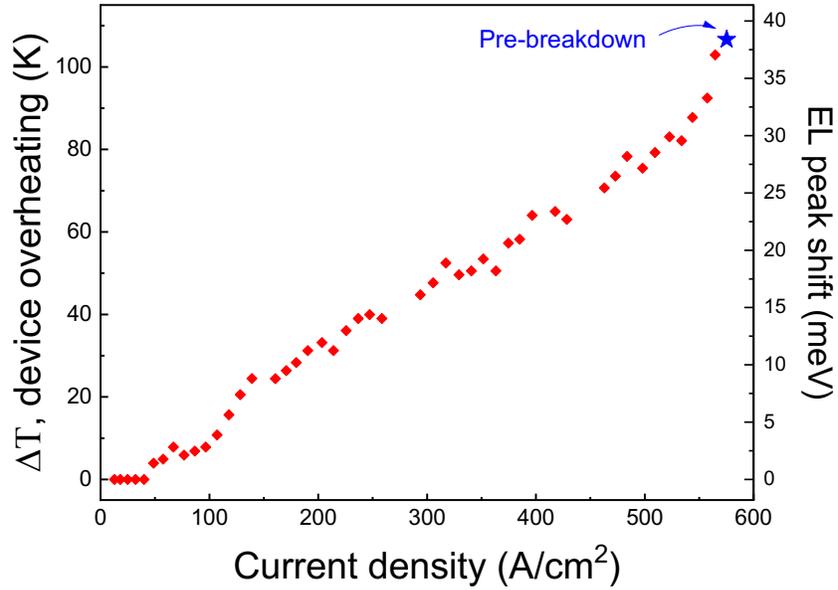

**Supplementary Figure 14. Device temperature change under electrical pumping.**

The increase in the device temperature ($\Delta T$) as a function of $j$. $\Delta T$ was derived from the shift of the 1S EL peak in the spectra shown in Supplementary Fig. 8b using $dE_g/dT = 0.365$ meV K$^{-1}$. At the highest $j$ (right before device breakdown), overheating of the active ccg-QD layer exceeds 100 K.